# A numerical tool for efficient analysis and optimization of offshore wind turbine jacket substructure considering realistic boundary and loading conditions

Zhenyu Wang, Selase Kwame Mantey, Xin Zhang*

*College of Civil Engineering and Architecture, Zhejiang University, 310058, China*

*Corresponding author. Xin Zhang, E-mail address: xin.zhang1259@zju.edu.cn*

**Abstract**

The jacket substructure is a critical component of the offshore wind turbine (OWT) that is the interface between the transition piece at the top and the grouted connection. This paper presents a comprehensive study on the optimization of a jacket substructure to achieve greater cost efficiency while maintain acceptable structural performance. A fast parametric finite element modelling (FEM) approach for jacket substructures was firstly proposed. The generated models took into account realistic loading conditions, including self-weight, wind load and section-dependent wave load, and soil-pile interaction. Parametric studies were conducted afterwards to investigate the trends of the mass and response of the jacket substructure with respect to the variation of geometric and sectional parameters. Optimizations of the jacket substructure were carried out using parametric optimization and numerical genetic algorithm (GA) optimization under three different optimization strategies corresponding to three groups of objective and constraint functions. The trends obtained by parametric analysis were used to guide the parameter selection in parametric optimization, while a rank-based mutation GA was established with the proposed efficient FEM embedded in as the solver to the optimization objective and constraint functions. Parametric optimization gained its advantage in computational efficiency, and the mass reduction were 6.2%, 10% and 14.8% for the three strategies respectively. GA optimization was more aggressive as the mass reductions were 16.8%, 22.3% and 34.3% for the three strategies, but was relatively more computational intense. The two proposed optimization methods and the three optimization strategies are both expected to be applied in practical engineering design of OWT jacket substructure with good optimization output and high computational efficiency.

# 1. Introduction

Renewable energy is experiencing rapid growth and is projected to contribute to approximately 95% of the overall increase in global power capacity by the year 2026 [1]. Among all renewable energy resources, the wind energy is believed to be one of the most promising types [2–4], The offshore environment has vast availability of sea area for their installation and the sea surface offers steady wind speed and uniform topology [4].

The jacket substructure was adopted based on the initial design of the offshore oil and gas platforms [5], and jacket substructure foundations are efficient for water depth of 30-80 m [5]. Given the potential exposure to more extreme environmental conditions and greater susceptibility to severe loads, the design analysis of wind turbine jacket structures necessitates a high level of scrutiny and attention. A more comprehensive and meticulous examination is required to ensure optimal design performance.

The lattice of the jacket substructure is made of the main chord members or leg members and the internal bracing members. The main function of the bracing members is to resist the deformation of the jacket [6]. Among the various bracing configurations available, such as the X-brace, Z-brace, and K-brace [6–9], research has indicated that the X-braced jacket substructure offers the most advantageous properties, demonstrating superior static/dynamic stiffness in all directions, nonetheless has more weld joints and higher mass [5–7]. Therefore, this paper will take the X-braced jacket substructure as the research object to study its structural response under multiple loads, and to find the optimized structure configurations.

The optimization of the jacket substructure will help to reduce its cost, increase the structural stiffness hence improve the overall performance of the structure. The jacket substructure is essentially a truss structure, and the optimization of a truss structure can be classified into three categories, namely, topology, shape and sized. Topology optimization seeks to find the optimal joints configuration, shape optimization finds the optimal coordinates of the joints in the design space, and size optimization deals with the selection of the optimal cross-section areas for the elements of the structure [10–12]. Zhang et al [13], proposed a feasible topology optimization method of the jacket structure at the conceptual design phase. Similarly, Tian et al [8] used topology optimization to reduce the jacket mass by 13.7% [12], the equivalent maximum stress reduced by 46.31 %, and the ultimate capacity of the optimized structure was higher than the pre-optimized structure [12]. Due to the predefined configuration of jacket structure of either 3-leg or 4-leg, there is not any major work on the shape optimization.

Size optimization is one of the most common types of structural optimization applied in industry [14]. It searches for the minimum weight of the truss structures subject to multiple loading conditions under the constraints on the member stresses and displacements [14,15]. Among past research works, Motlagh et al [16] optimized the diameter and thickness of an offshore oil platform jacket, which has different load response compared to the offshore wind turbine (OWT) structure, using the genetic algorithm (GA), and a weight reduction of 15% was achieved under the ultimate load condition [16]. Sandal et al [17], only considering the wind environmental load, geometrically optimized the jacket substructure by focusing on the diameter and thickness of the jacket members, under fatigue, ultimate limit state and frequency constraints. The optimization achieved 10% reduction in mass. They found that the base width is a sensitive factor that influences the optimization process [17]. In addition, thy emphasized the effect of the soil characteristic on the optimization results [17]. Oest et al [18] assumed fixed boundary condition with the foundation grouted connections. The jacket was optimized with respect to member diameter and thickness with 20 design variables and achieved a 40% mass reduction using sequential linear programming [18], under both fatigue and ultimate limit state constraints. The displacement response on the optimized model was 80 mm higher than the original model [18].

It can be concluded from the results of the listed research works that size optimizations on the geometric properties of jacket structural members (especially the diameter and thickness) can achieve favourable mass reduction and displacement control. However, the optimizations carried out did not consider the detailed geometric configuration of jacket substructures, whilst some considered the diameter and thickness, and others considered the base width only. Additionally, a realistic determination of the wave load magnitude, which is contingent upon the geometry and dimensions of the structural members, as well as their elevation relative to sea level, was not performed for some works. Furthermore, the potential movement of pile foundations was disregarded, with a fixed boundary condition being commonly assumed. Therefore, to optimize the jacket substructure more comprehensively with modelling the real load and boundary conditions, a more powerful but efficient modelling and optimization system is highly demanded.

This paper aims to introduce an efficient parametric modelling approach for OWT jacket structure with realistic boundary and loading conditions, which enables the execution of parametric and numerical optimizations with notable efficiency. The workflow of this work is shown in Fig 1. The paper is organized as follows. Section 2 introduces the load to be applied to the jacket structure model with emphasis on the wave load that may vary with water level, location and diameter of structural members. The implementation of applying loads is described as well. Section 3 presents the specifics of the modelling approach, and introduces the method for modelling soil-pile interactions. Parametric analysis is conducted in Section.4 to study the trends of mass and structural responses with respect to the variation of geometric and sectional parameters. With the trends obtained, parametric optimization is carried out in Section.5, and the numerical rank-based mutation GA optimization is conducted as well in Section.5. At the end of paper, main conclusions are drawn in Section 6.

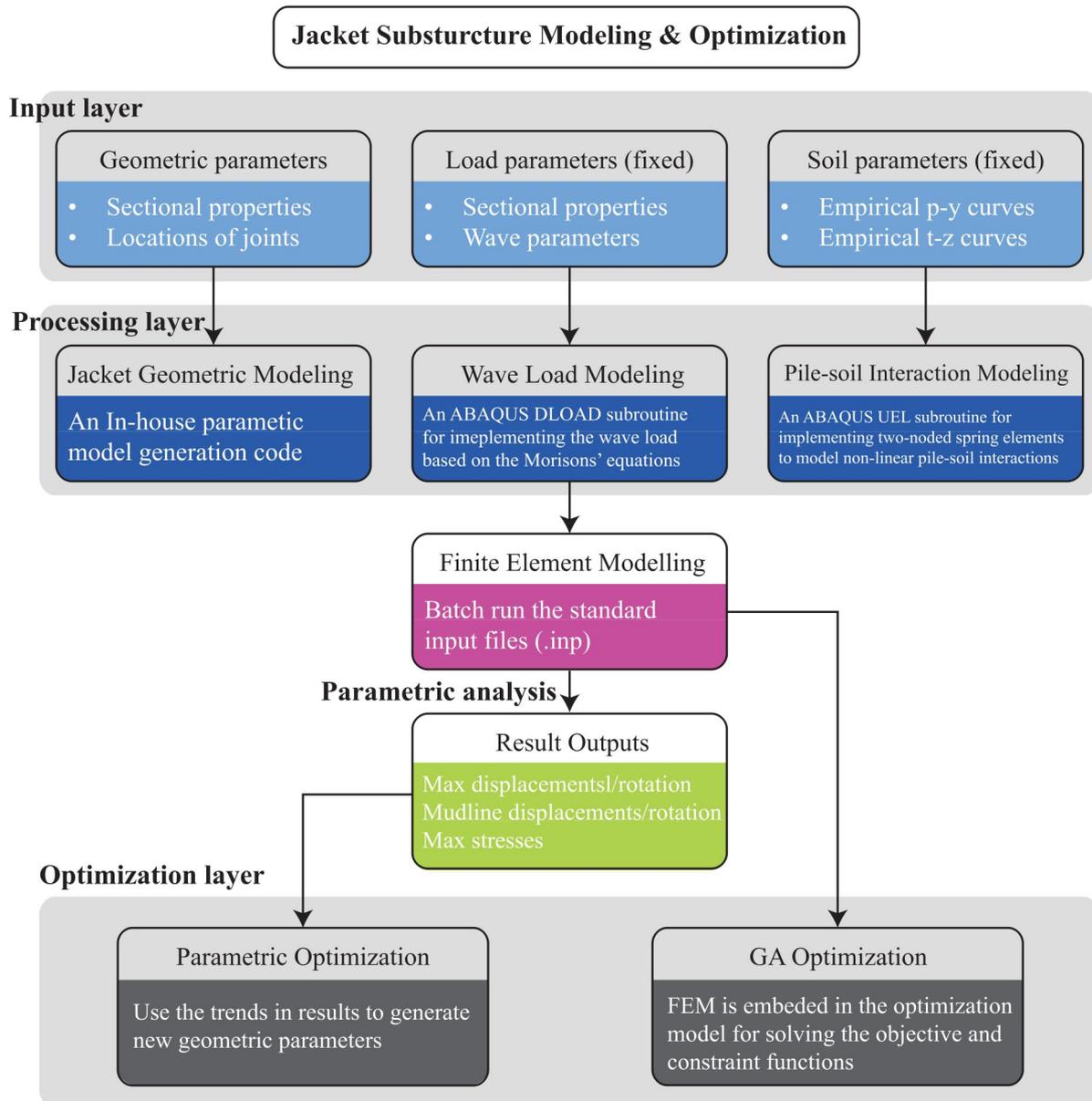

Fig 1. Flowcharts of optimization processes

## 2. The jacket loading

For the ultimate limit state design, the environmental loads from the wind and waves are the main load components applied to OWT jacket substructure. Additionally, the self-weight of the jackets and the dead loads from the rotor nacelle assembly (RNA), the tower and transition piece are accountable for the approximate ultimate load effect on the jacket substructure. Shittu et al, stated in their research that for the ultimate load design, the wind load is the design driving factor and the hydrodynamic load effects are secondary [19]. For design limit state of the jacket substructure under environmental loading, the key design variables are the wind speed, significant wave height and peak wave period. In this study, all these load types are considered in the analysis and optimizations.

### 2.1 Wind and Dead Loads

The ultimate load on top of the jacket structure applied at the reference point (RP, as shown in Fig 2) is made of the estimated wind load and the self-weight of the RNA and tower. From Chao et al [20], the comparative study of the three load simulation approaches (fully coupled, sequentially coupled, and uncoupled), the different load simulation approaches have little effect on the mean values of the loading and displacement. The values of unfactored loads from wind, RNA, and tower are shown in Table 1. The unfactored loads were obtain from a design research institute in China.

Table 1. Ultimate loads applied on the jacket from wind and top self-weight

| Load case | Fz (kN) | Fx (kN) | Fy (kN) | Mz (kNm) | Mx (kNm) | My (kNm) |
|---|---|---|---|---|---|---|
| Extreme operating condition (wind1) | - | 5071 | 5071 | -33324 | 423875 | 423875 |
| Normal operating condition (wind2) | - | 2104 | 2104 | -24993 | 216729 | 216729 |
| Dead load (DL) | -14972 | | | | | |

### 2.2 Wave Load

The wave model used in the developed model was based on Lars Skjelbreia and other related researches [21–25]. The model depends on period ($T_p$), significant wave-height ($H_s$), water depth ($z_s$), and wave-length ($L$). The fifty year and five-year wave parameters used in the calculations are listed in Table 2.

Table 2. Fifty years and five years wave parameters

| Return period | $H_s$ (m) | $T_p$ (s) | $L$ (m) |
|---|---|---|---|
| 50 years (wave1) | 13.98 | 18.06 | 308.07 |
| 5 years (wave2) | 7.73 | 12.33 | 161.26 |

The wave load was calculated based on the Morisons' equations, and its horizontal and normal components are defined as:

$$f_H(t) = \rho C_M A a + \frac{1}{2}\rho C_D D_v |v|v \tag{1}$$

$$f_N(t) = \rho C_M A a \sin(\alpha) + \frac{1}{2}\rho C_D D_v |v|v \sin(\alpha) \tag{2}$$

where $C_D = 1.2$ is the initial drag force coefficient, and $C_M = 2.0$ is the mass coefficient. These two coefficients were further modified based on the water depth and the inclination angles of structural members with respect to the wave propagation direction (as shown in Fig 2). For members inclined at an angle ($\alpha$), the force is decomposed into the normal force (Eq.) on the member and the lift force along the member. $v$ is the fluid particle velocity, $a$ is the fluid particle acceleration, $A$ is the cross-sectional area, $D_v$ is the diameter of pipe, and $\rho$ is the density of fluid. The only two unknown but crucial variables, $v$ and $a$, are obtained by:

$$v = -\bar{c}\epsilon_1 \tag{3}$$

$$a = \frac{2\bar{c}\pi\epsilon_3}{T_p} \tag{4}$$

where $\bar{c}$ is the wave celerity, and it is formulated as:

$$\bar{c} = \sqrt{\frac{\tanh(\beta z_s)}{\beta} g \left(1 + \lambda^2 C_1 + \lambda^4 C_2\right)} \tag{5}$$

in which $\beta = 2\pi/L$, $g$ is the gravity acceleration. $C_1$ and $C_2$ are dimensionless coefficients related to $\beta$ and $z_s$. The coefficient $\lambda$ is determined through numerically solving the following equation:

$$-\frac{H_s}{2}\beta + \lambda + B_{33}\lambda^3 + (B_{35} + B_{55})\lambda^5 = 0 \tag{6}$$

and the two parameters $\epsilon_1$ and $\epsilon_3$ are determined using the Stoke's 5th theory:

$$\epsilon_1(t) = \sum_{n=1}^{5} n\kappa_n \cosh(n\beta(z_s+z))\cos(n(\beta x - \omega t)) \tag{7}$$

$$\epsilon_3(t) = \sum_{n=1}^{5} n^2\kappa_n \cosh(n\beta(z_s+z))\sin(n(\beta x - \omega t)) \tag{8}$$

where $\kappa_n$ are coefficients introduced for convenience. They are defined as polynomials with respect to $\lambda$, where the coefficients of the polynomials are denoted as $A_{ij}$. $x$, $\omega$ and $t$ indicate the wave location, frequency and the time. Coefficients $A_{ij}$, $B_{ij}$ and $C_i$ are functions of the ratio of water depth to wave length, and they can be found in the stated reference [26].

In the developed finite element model of this paper, the wave load is applied on each integration point as a distributed load. It can be found in Eqs. to that the applied wave load depends on the diameter ($D_v$), inclination angle ($\alpha$), and the elevation ($z$) of the structural member, which means that the magnitude of load would be different on each integration point. Therefore, we adopted the DLOAD subroutine to implement such variation, and the pseudocode for the subroutine is shown below.

| Algorithm 1: Subroutine DLOAD | |
|---|---|
| INPUT:     *coords*(3) | coordinates of the integration point |
|              eleSectList.txt | sectional properties |
|              geomList.txt | orientation of structural members |
|              $T_p$, $H_s$, $z_s$, $L$ | |
| OUTPUT:    $F$ | magnitude of the load |

1. **Read** *geomList.txt*, locate the integration point on a specific member;
2. Calculate $\sin(\alpha)$ based on the orientation of the member;
3. **Call subroutine CsModi()** to modify $C_D$ and $C_M$ based on the orientation of the member;
4. **Read** *eleSectList.txt*, find the sectional properties of the member;
5. Calculate coefficients $A_{ij}$, $B_{ij}$ and $C_i$, and further $\overline{c}$;
6. Solve $\lambda$ using the Newton's method;
7. Build $\epsilon_1$ and $\epsilon_3$ as the functions of $t$;
8. Calculate $f_N(t)$, and $F = \max(f_N(t))$

### 2.3 Design Load Combinations

The design load combination was estimated by combining the load cases in Table 1 and Table 2 based on the DNV code J101 and Chinese code GB 50009-2012 [27] recommended load combination. The load combination with the worst-case ultimate response is 1.5*(wind1) + 1.35*0.7*(wave1) + 1.0* DL.

## 3. Finite element modelling

### 3.1 The Jacket Substructure Model

A total of 196 components were modelled and divided into 26 groups according to different sectional properties as shown in Fig 2, and the original geometric properties of the jacket substructure are outlined in Table 3. In the absence of the transition piece, the top of the jacket was modelled using a structural representation comprising a 50 mm thick 2D plate and four cylindrical pipes with 1550 mm in diameter and 200 mm in thickness. The dimensions of the structural representations were selected to confer sufficient structural rigidity for the effective transmission of the applied load.

The horizontal, vertical, bending, and torsional loads, from wind and the self-weight of the wind turbine, were applied at the top of the jacket, and the wave load was applied simultaneously on the members below the water level. Preliminary analysis of the jacket structure encompassed evaluations across three different water depths: low water level (LWL, 62-3.04=58.96 m), mean water level (MWL, 62 m), and high water level (HWL, 62+4.39=66.39 m). The corresponding simulated maximum displacements for these conditions were 81.36 mm, 82.20 mm, and 83.88 mm, respectively; the maximum displacements at the mudline level were 16.55 mm, 17.05 mm, and 17.56 mm, respectively; and the highest stress identified for each water level manifested as 290.60 MPa, 288.21 MPa, and 288.70 MPa, sequentially. Based on the results obtained, the HWL was selected for subsequent parametric studies and the ensuing optimization endeavours.

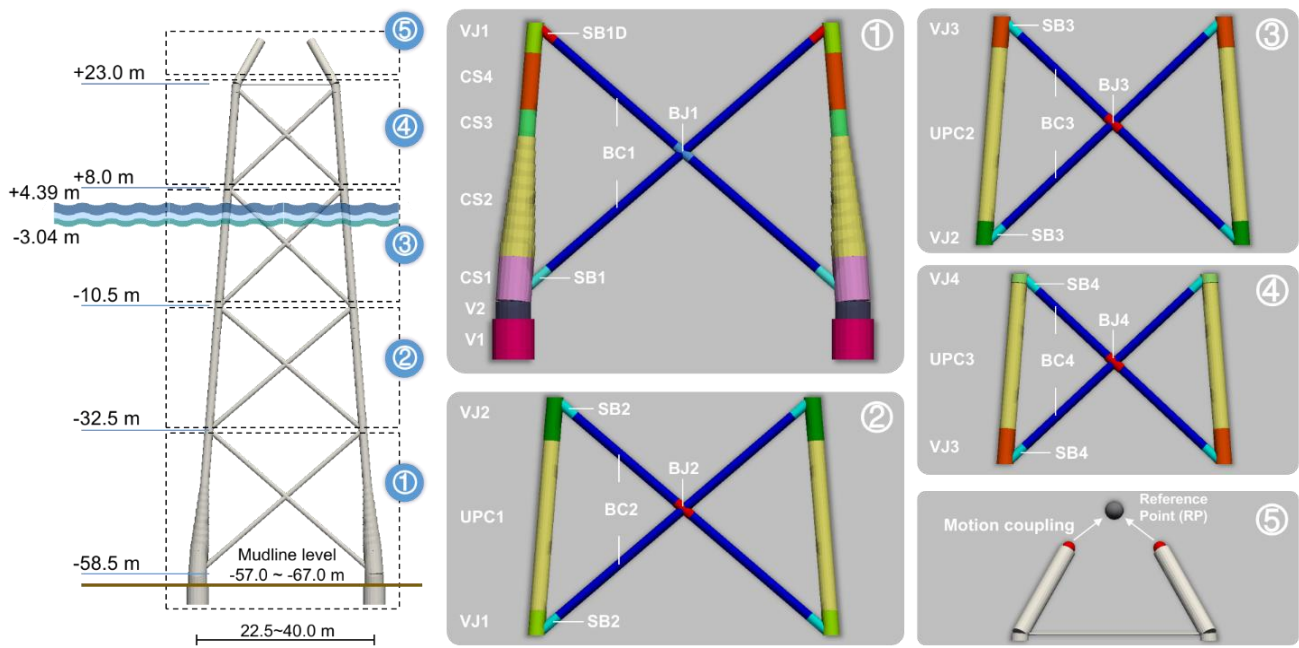

Fig 2. Labelling of structural members

Table 3. Geometric and sectional properties of the original jacket substructure

| Geometric properties | | | | |
| --- | --- | --- | --- | --- |
| Label | Diameter | Thickness | Length (mm) | Remarks |
| V1 | 4200 | 65 | 3000 | Pile above mudline |
| V2 | 3600 | 85 | 2930 | Jacket leg above grout connection |
| CS1 | 3600 | 85 | | Level-1 |
| CS2 | 3600-1450 | 85 | 11930 | Level-1 |
| CS3 | 1450 | 75 | 3000 | Level-1 |
| CS4 | 1450 | 75 | 4778 | Level-1 |
| VJ1 | 1460 | 80 | 2991 | Joint level-1 to level-2 |
| VJ2 | 1450 | 75 | 3271 | Joint level-2 to level-3 |
| VJ3 | 1450 | 75 | 2614 | Joint level-3 to level-4 |
| VJ4 | 1470 | 85 | 1648 | Level-4 |
| UPC | 1420 | 60 | 18858 | Level-2 |
| UPC2 | 1420 | 60 | 15670 | Level-3 |
| UPC3 | 1420 | 60 | 12023 | Level-4 |
| SB1 | 975 | 60 | 4432 | Level-1 |
| SB1D | 850 | 55 | 2886 | Level-1 |
| SB2 | 810 | 45 | 3112-2915 | Level-2 |
| SB3 | 740 | 45 | 2989-2780 | Level-3 |
| SB4 | 640 | 40 | 2989-2780 | Level-4 |
| BC1 | 850 | 55 | 15452, 14446, 15378, 14372 | Level-1 |
| BC2 | 800 | 40 | 14354, 13348, 11966, 10960 | Level-2 |
| BC3 | 710 | 30 | 11799, 10837, 9847, 8885 | Level-3 |
| BC4 | 620 | 30 | 9333, 8371, 7826, 6864 | Level-4 |
| BJ1 | 860 | 60 | 2012 | Level-1 |
| BJ2 | 810 | 45 | 2012 | Level-2 |
| BJ3 | 730 | 45 | 1924 | Level-3 |
| BJ4 | 630 | 35 | 1924 | Level-4 |
| Bottom width | 35000 | | | |
| Top width | 16000 | | | |

An efficient model generator was developed to expedite the creation of finite element models for jacket substructures. It enables the rapid generation of a batch of such models within a remarkably short time period (approximately 0.1 seconds per model, taking into consideration of our PC configuration). Each model necessitates the utilization of two standalone scripts, denoted as *.jct* and *.sec* files, which are employed to pass in geometric and sectional parameters, respectively. The user has the flexibility to construct these scripts manually by inputting values or, alternatively, to employ a batch generator for their automated generation. The Gmsh Python API [28] was adopted to draw the model's configuration and execute the meshing process. The resulting mesh file adhered to the Abaqus scripting file format (.inp), and it can be incorporated into a pre-established main script. This main script encompassed all the essential information needed for finite element modelling, including specifics related to the DLOAD and UEL. A mesh sensitivity analysis was performed with characteristic element sizes of 650 mm, 675 mm, 750 mm, 1000 mm and 1500 mm, by using the maximum stress as the index, the simulation results converged at the size of 1000 mm. Full model of the jacket substructure is shown in Fig 3.

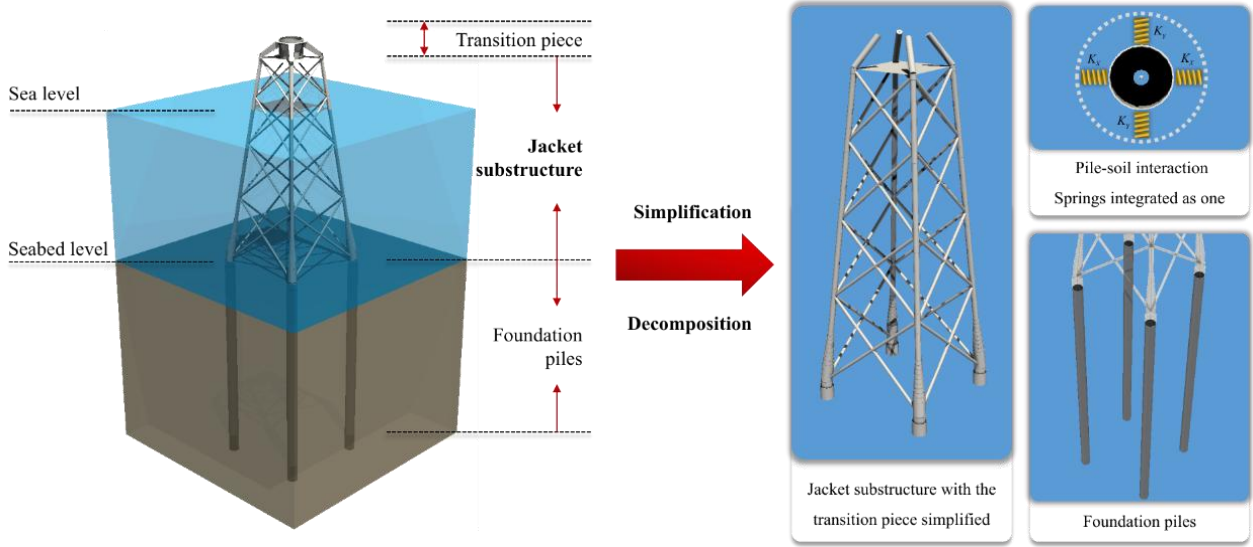

Fig 3. The full jacket model

**3.2 Modelling of the nonlinear pile-soil interaction**

The nonlinear foundation pile-soil interaction was modelled using two-noded spring elements. The two nodes of a spring element were initially overlapped, and one of the nodes is on the pile while the other one is fixed throughout analysis. The governing equation for the entire model writes:

$$\int_{\Omega_0} \boldsymbol{\sigma} : \delta^{sym}\boldsymbol{\varepsilon} \mathrm{d}V - \int_{S_{int}} \mathbf{T} \cdot \delta \boldsymbol{\Delta}_{int} \mathrm{d}S = \int_{\Gamma_s} \mathbf{F}_w \cdot \delta \mathbf{u}_w \mathrm{d}\Gamma + \mathbf{F}_{ref} \cdot \delta \mathbf{u}_{ref} \tag{9}$$

where $\boldsymbol{\sigma}$ and $\boldsymbol{\varepsilon}$ are the stress and strain tensors of the jacket structure and piles ($\Omega_0$). $\mathbf{F}_w$ is the distributed wave load vector applied on jacket structural members under the sea level and above the mudline level ($\Gamma_s$), and $\mathbf{u}_w$ is the corresponding displacement vector due to wave load. $\mathbf{F}_{ref}$ is the concentrated force vector applied on the reference point, and $\mathbf{u}_{ref}$ is the displacement vector of the reference point. $\mathbf{T}$ is the distributed pile-soil interaction force vector including two components in the lateral directions ($p_{H1}$ and $p_{H2}$ due to soil resistance) and one in the vertical direction ($t_V$ due to vertical shearing), and $\boldsymbol{\Delta}_{int} = [\Delta_{H1} \quad \Delta_{H2} \quad \Delta_V]$ is the relative displacement vector of the interface.

Through classical finite element discretization of the governing equation, the stiffness matrix of a spring element is derived as:

$$\mathbf{K}_e = L_e \mathbf{B}^T \frac{\partial \mathbf{T}}{\partial \boldsymbol{\Delta}_{int}} \mathbf{B} \tag{10}$$

and the element nodal force vector is:

$$\mathbf{f}_e = L_e \mathbf{B}^T \mathbf{T}(\boldsymbol{\Delta}_{int}) \tag{11}$$

where $L_e$ is the representative length of an interface segment that a spring element is responsible for. **B** is the operator relating the spring element's nodal degree of freedoms ( $\mathbf{u}_s = [u_{1x}\ u_{1y}\ u_{1z}\ u_{2x}\ u_{2y}\ u_{2z}]$ ) and the relative displacement of the interface:

$$\mathbf{B} = \frac{\partial \mathbf{\Delta}_{int}}{\partial \mathbf{u}_s} = [\mathbf{I}\ -\mathbf{I}] \tag{12}$$

with **I** a 3×3 identity matrix.

The expression of the interaction force as the functions of relative displacements are determined empirically, and the *p-y* (for $p_H - \Delta_H$) and the *t-z* (for $\tau_V - \Delta_V$) curves in the DNV code [29,30] were used. The soil strata were divided into four layers (silt, muddy silt clay, silty clay, and silty sand) corresponding to four different p-y and t-z curves, and each individual curve was divided into a three-continuous piecewise function. The empirical p-y and t-z curves are plotted in Fig 4 and Fig 5. Noting that $p_H$ and $\Delta_H$ are the resultant horizontal interaction force and relative displacement, defined as:

$$p_{H1} = \frac{\Delta_{H1}}{\Delta_H} p_H, \quad p_{H2} = \frac{\Delta_{H2}}{\Delta_H} p_H \quad \text{with} \quad \Delta_H = \sqrt{\Delta_{H1}^2 + \Delta_{H2}^2} \tag{13}$$

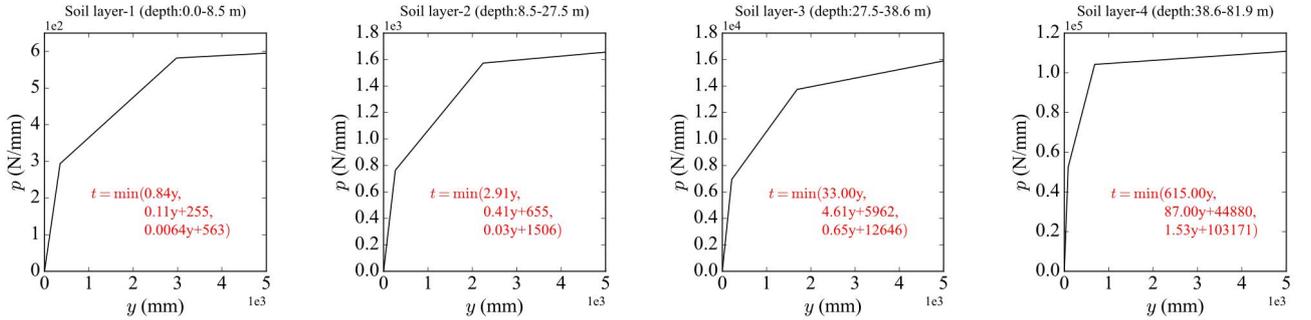

Fig 4. The jacket pile spring support model *p-y* curves

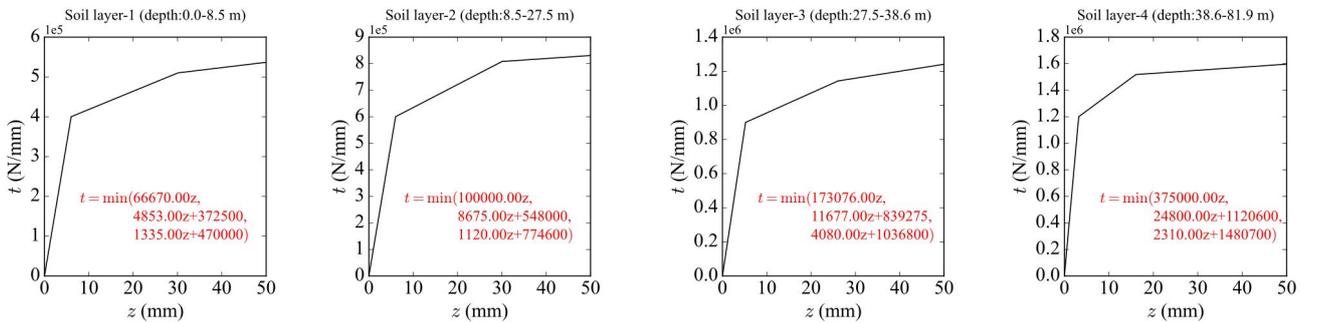

Fig 5. The jacket pile spring support model *t-z* curves

The two-noded spring element was implemented using the user-defined-element (UEL) subroutine in ABAQUS. Jacket models with fixed support and incorporating pile-soil interactions were simulated and compared. The comparatively analysis was aimed at elucidating the substantial impact that pile-soil interactions on the responses of offshore wind turbines. Table 4 outlined the results of comparison. The change in the maximum stress is marginal, from 288.23 MPa to 288.70 MPa as shown in Fig 6. The increase in displacement and rotation was prominent both at the top of the jacket and at the mudline level.

Table 4. Comparison between fixed support and spring support

| Description | Fixed Support | Spring Support | Difference % |
| --- | --- | --- | --- |
| Maximum top stress (MPa) | 288.23 | 288.70 | 0.16 |
| Maximum top displacement (mm) | 63.75 | 87.15 | 36.71 |
| Max rotation (°) | 0.2946 | 0.3088 | 4.82 |
| Displacement at mudline (mm) | 0 | 19.04 | 100.00 |
| Rotation at mudline (°) | 0 | 0.1203 | 100.00 |

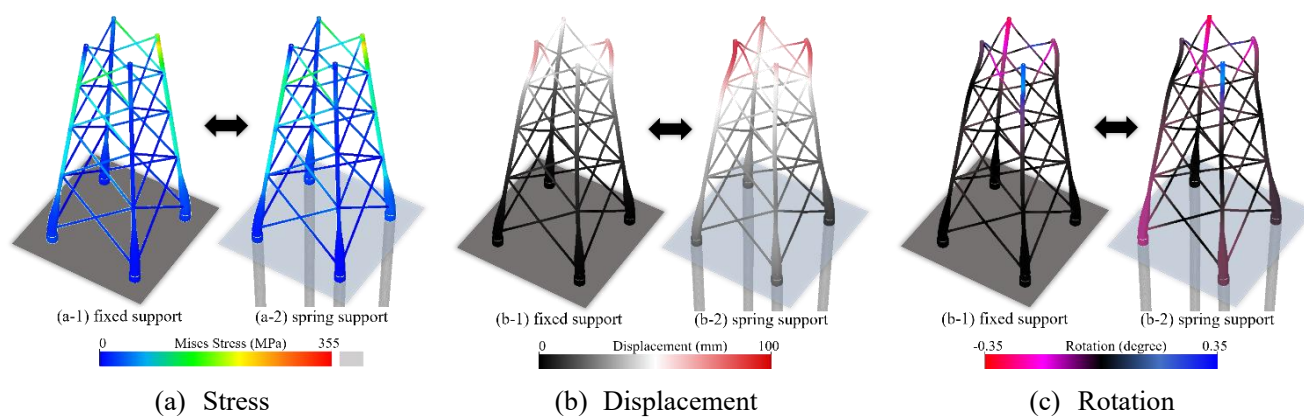

(a) Stress  (b) Displacement  (c) Rotation

Fig 6. Comparison between fixed support and spring support

## 4. Parametric Analysis

To develop an effective model for parametric optimization, a series of parametric studies were conducted. These studies aimed to explore the relationship between rotation, displacement, and stress in response to variations in the base width, diameter, and thickness parameters. Specifically, the change in pile length (PL) above mudline level, base width (BW), brace diameter (BD), brace thickness (BT), leg diameter (LD), and leg thickness (LT) were examined.

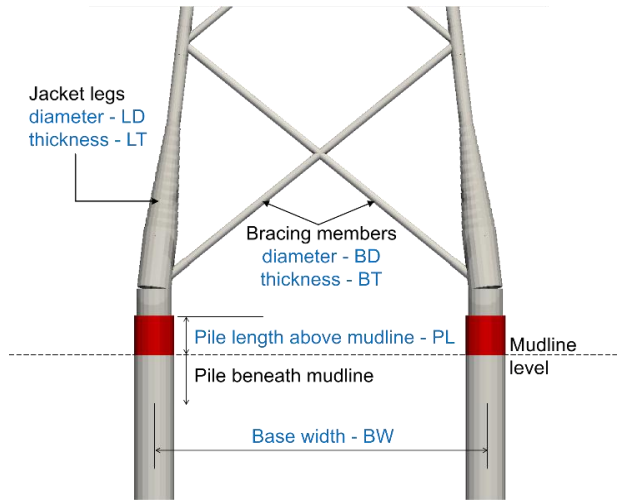

Fig 7. Illustration of parameters to be used in the analysis

**4.1 Change in pile length above mudline**

The pile length above the mudline level (PL), as depicted in Fig 7, was varied from 1400 mm to 5900 mm to investigate the influence of PL on the response of the jacket substructure. As shown in Fig. 8, it was found that augmenting the PL resulted in a marginal increase of merely 0.27% (+0.8 MPa) in the maximum stress. Also, the maximum displacement, both overall and at the mudline level, exhibited an upward trend (as shown in Fig 8(b)). Notably within the varying range, the maximum overall displacement escalated by 9.4% (from 84.04 to 91.90 mm), and the maximum pile displacement at the mudline level increased by 10.5% (from 18.37 to 20.30 mm).

In addition, the maximum rotation at the jacket top and the pile rotation at the mudline level also increase by 3.59% (marginally from 0.362° to 0.375°) and 34.16% (significantly from 0.120° to 0.161°) respectively, as shown in Fig 8(c). The above analysis indicates that the impact of increasing PL is pronounced on the overall displacement and the pile rotation at the mudline level. The mass of the jacket structure was not affected by the PL variation.

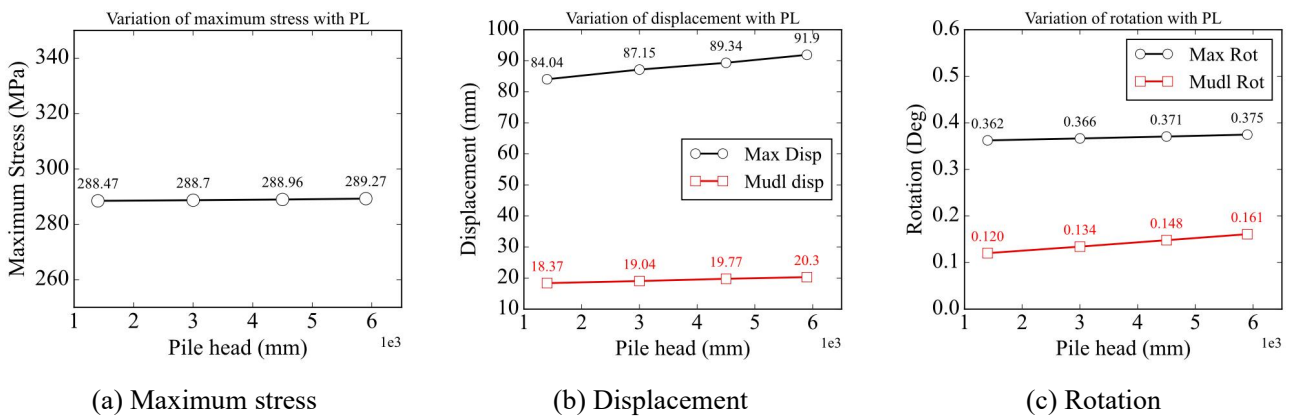

(a) Maximum stress  (b) Displacement  (c) Rotation

Fig 8. Change in the pile length above mudline

## 4.2 Variation of base width

The base width (BW as illustrated in Fig 7) was varied within the range of 22500 mm to 40000 mm to study its impact on the response of the jacket substructure. As evidenced in Fig 9(a), the increase in BW led to a substantial 13.4% reduction in the maximum stress (from 320.31 to 277.43 MPa). As evident from Fig 9 (b) and (c), the variations in BW exerted more significant impact on response of the overall jacket structure compared to that of the piles at the mudline level both in terms of the maximum displacement and the rotation. With the increase in BW, the maximum overall displacement decreased by 47.9% (from 146.5 to 76.34 mm), while the reduction in the maximum pile displacement at the mudline level was 8.5% (from 21.25 to 19.44 mm, but the curve did not show a monotonous downward trend). The maximum overall rotation occurred at the top of the jacket members decrease by 0.063° while the maximum pile rotation at the mudline level increased by 0.025°. Although increasing BW can efficiently reduce the stress and displacement of the structure, it also greatly increased the mass of the structure. Within the predefined varying range, the mass of the jacket substructure increased from 1664 to 1838 ton. The variation of stress, displacement and rotation per unit mass increment was -0.246 MPa/ton, -0.403 mm/ton (overall), -0.010 mm/ton (mudline level), -3.62e-4°/ton (overall) and +1.44e-3°/ton (mudline level).

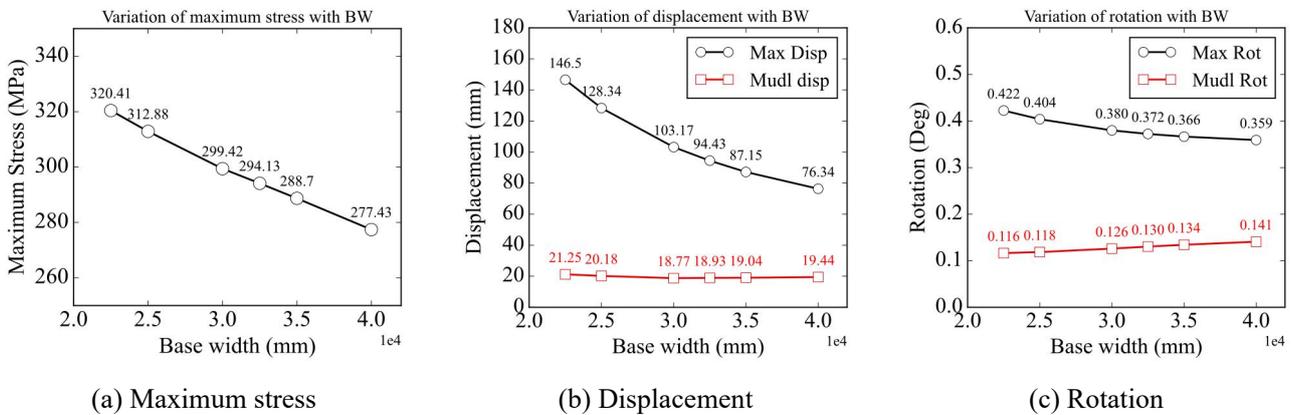

(a) Maximum stress     (b) Displacement     (c) Rotation

Fig 9. Change in base width

## 4.3 Variation of diameter of bracing members

The sensitivity of diameter of bracing members (BD) on the jacket structural response was analysed in this section with BD varied in combinations (Comb.1 to 6) as outlined in Fig 10.

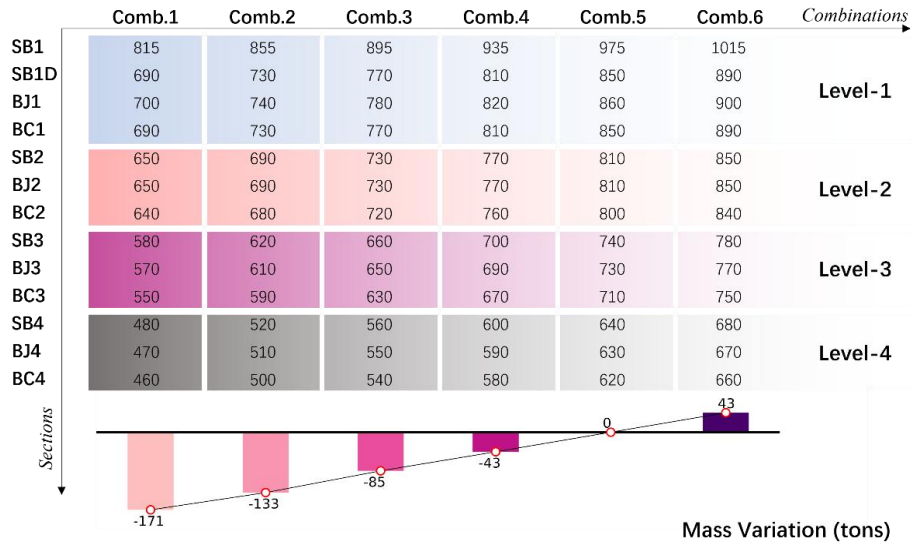

Fig 10. Variation of brace diameters

The diameter of the bracing member BC4 was used as the reference parameter. As shown in Fig 11(a), it was found that the increase in the BD corresponded to a marginal decrease in the maximum overall stress by 0.8% (from 290.17 to 287.75 MPa). The maximum overall displacement and the maximum pile displacement at the mudline level (Fig 11 (b)) decreased with the increasing of BD by 7.3% (92.92 to 86.62 mm) and 9.7% (20.65 to 18.65 mm), respectively. In addition, the maximum overall rotation and the pile rotation at the mudline level reduced by 8.2% (0.392° to 0.360°) and 28.0% (0.176° to 0.120°) respectively (Fig 11 (c)). The variation of stress, displacement and rotation per unit mass increment was -0.011 MPa/ton, -0.032 mm/ton (overall), -0.009 mm/ton (mudline), -1.50e-4°/ton (overall) and -2.34e-4°/ton (mudline).

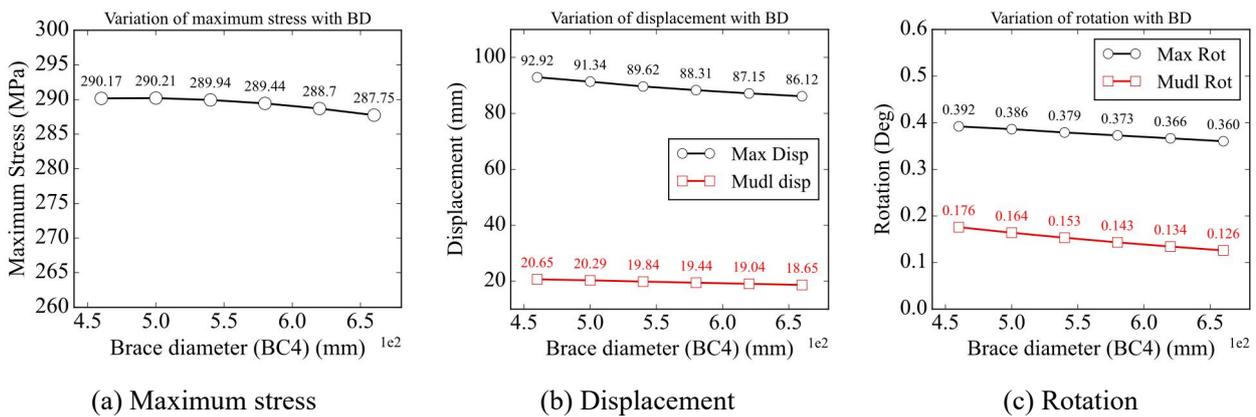

(a) Maximum stress     (b) Displacement     (c) Rotation

Fig 11. Change in brace diameter

**4.4 Variation of thickness of bracing members**

The influence of thickness of bracing members (BT) on the jacket structural responses was analysed in this section. For the 13 bracing, analysis on the 6 thickness combinations (as shown in Fig 12) were performed.

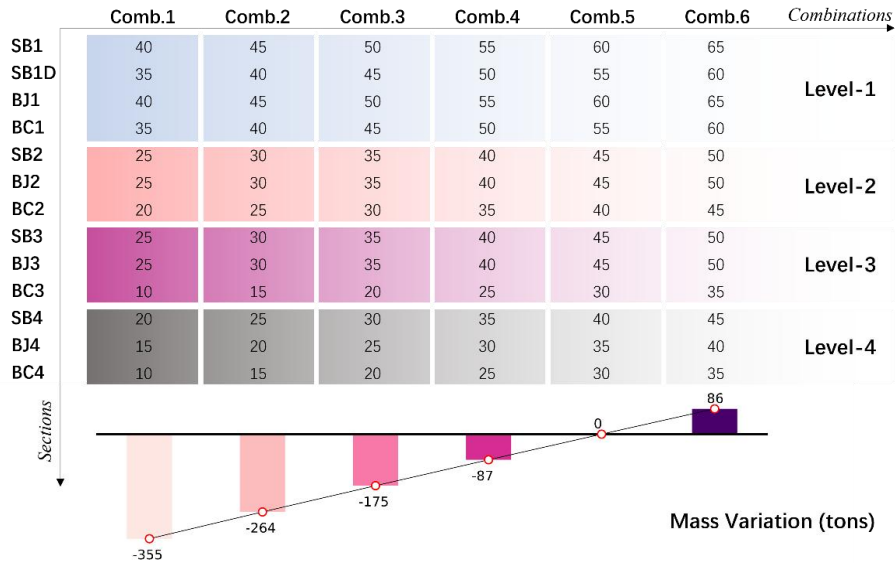

Fig 12. Variation of brace thickness

The member BC4 thickness was used as the reference parameter. As shown in Fig 13(a), the increase in BT corresponded to a significant decrease in the maximum stress, especially when the thickness of BC4 was no larger than 20 mm (from 374.00 to 291.72 MPa). Besides, the maximum overall displacement and the maximum pile displacement at the mudline level (Fig 13 (b)) decreased with the increasing of BT by 26.3% (116.79 to 86.10 mm) and 6.7% (20.28 to 18.93 mm) respectively. Similarly, more dramatic decrease in displacements were found when the thickness of BC4 was less than 20 mm. For the maximum rotation, decreases of 18.7% (0.443° to 0.360°) and 21.7% (0.166° to 0.130°) were achieved for the jacket structure and the pile at the mudline level respectively, as shown in Fig 13(c). The variation of stress, displacement and rotation per unit mass increment was -0.193 MPa/ton, -0.070 mm/ton (overall), -0.003 mm/ton (mudline level), -1.88e-4°/ton (overall) and -8.16e-5°/ton (mudline level).

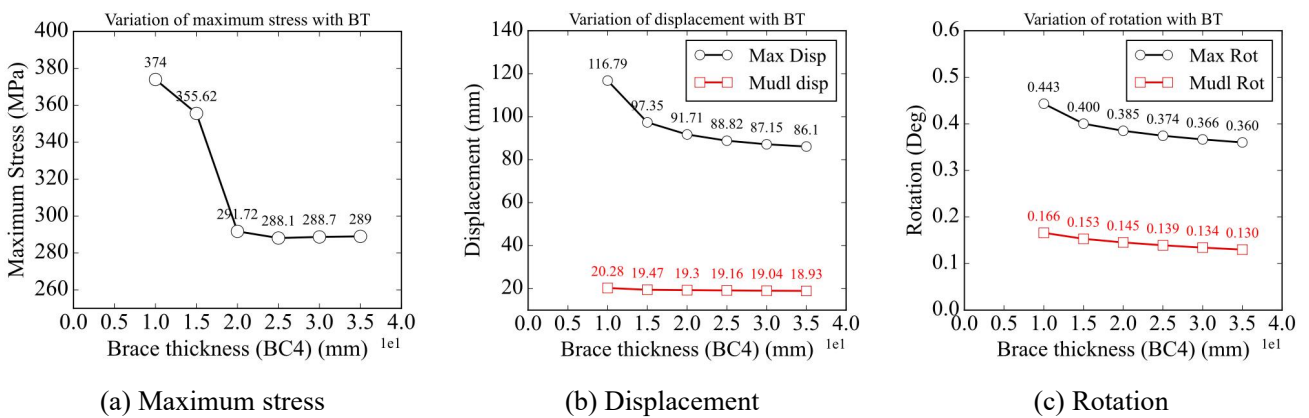

(a) Maximum stress  (b) Displacement  (c) Rotation

Fig 13. Change in braced thickness

## 4.5 Variation of diameter of main jacket legs

The diameters of main jacket legs (LD) were varied in combinations as outlined in Fig 14 to analysis their impact on the jacket structural responses.

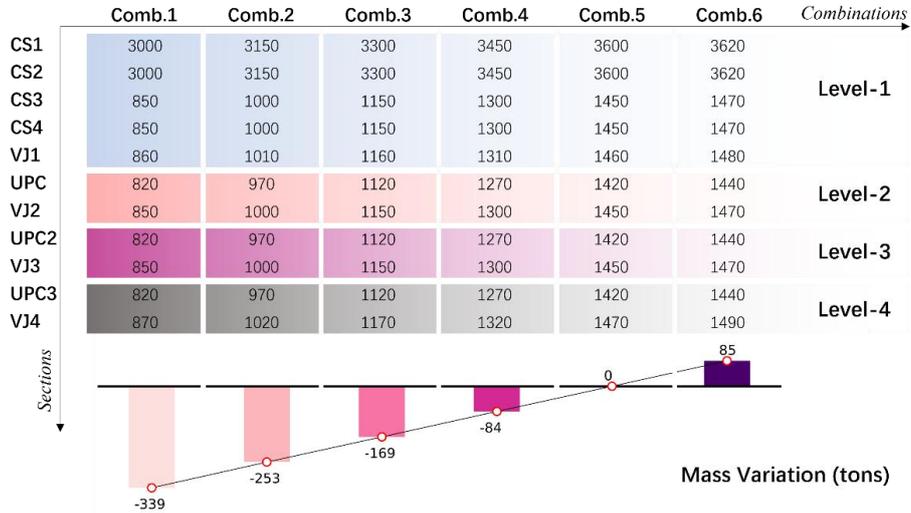

Fig 14. Variation of leg diameter

The diameter of the jacket legs UPC3 was used as the reference parameter. As shown in Fig 15(a), the scatter plot indicates clear trend where the maximum stress generally decreased with the raising of LD. The reduction in maximum stress amounted to approximately 23.5%, transitioning from 358.94 MPa to 274.47 MPa. As shown in Fig 15(b) and (c), the maximum displacement and rotation of both the jacket structure and the piles at the mudline level are approximately negative linear to LD. The maximum displacement decreased by 39.3% and 25.0% for the jacket structure and the piles at the mudline level respectively, and the reduction in the maximum rotation of the jacket structure was 26.1 %, while that of the piles at the mudline level was 8.5%. The variation of stress, displacement and rotation per unit mass increment was -0.199 MPa/ton, -0.124 mm/ton (overall), -0.014 mm/ton (mudline level), -2.90e-4°/ton (overall) and -2.83e-5°/ton (mudline level).

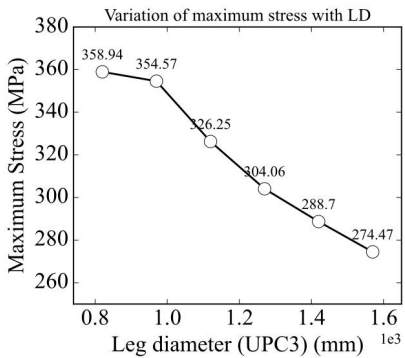

(a) Maximum stress

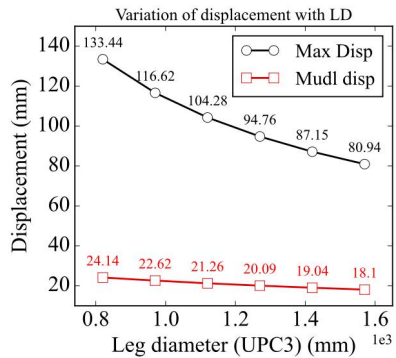

(b) Displacement

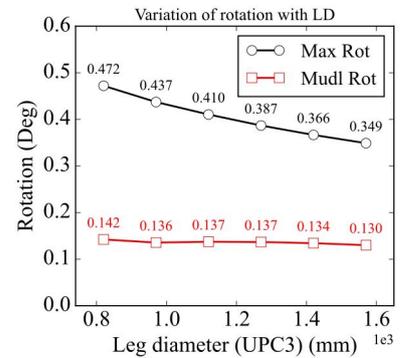

(c) Rotation

Fig 15. Change in leg diameter

## 4.6 Variation of thickness of main jacket legs

The thicknesses of the main jacket legs (LT) were varied according to the combinations outlined in Fig 16. The thickness of member UPC3 was used as the reference parameter for scatter plots.

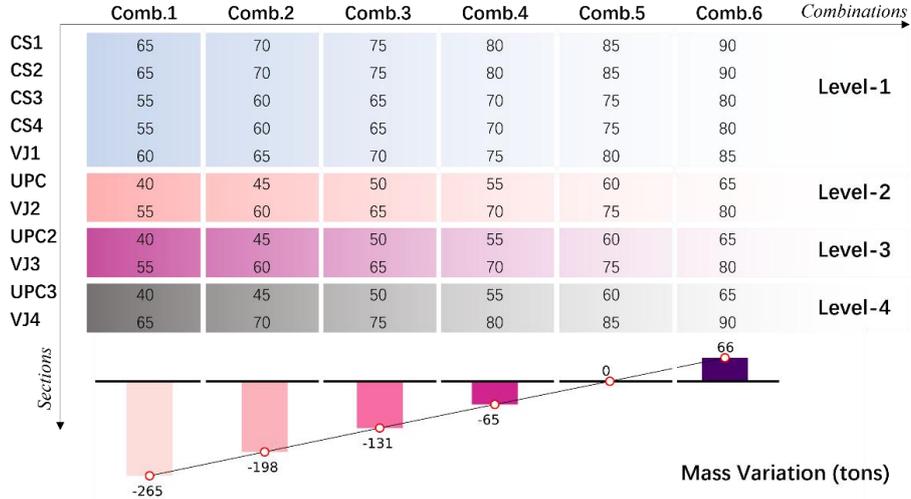

Fig 16. Variation of leg thickness

As shown in Fig 17(a), the increase in the LT corresponded to a significant decrease in the maximum stress by 24% (from 360.87 MPa to 274.18 MPa). Furthermore, as depicted in Fig 17(b) and (c), the fluctuations in LT had a more substantial impact on the overall jacket structure's response compared to that of the piles at the mudline level, encompassing both the maximum displacement and rotation. With the increase in LT, a noteworthy reduction of 24.5% was observed in the maximum overall displacement, declining from 110.62 mm to 83.53 mm. Concurrently, there was an 5.4% reduction in the maximum pile displacement at the mudline level, from 19.94 mm to 18.86 mm. Regarding rotation, the maximum overall rotation occurring at the top of the jacket members decreased by 0.048° (+11%), while the maximum pile rotation at the mudline level experienced a slight increase of 0.001°. The variation of stress, displacement and rotation per unit mass increment was -0.262 MPa/ton, -0.082 mm/ton (overall), -3.26e-3 mm/ton (mudline level), -1.50e-4°/ton (overall) and +0.30e-5°/ton (mudline level).

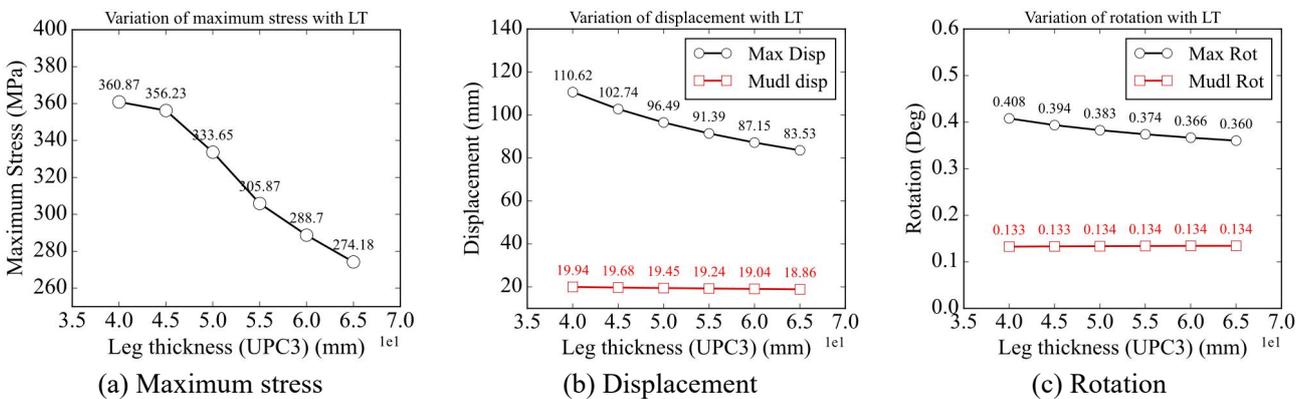

(a) Maximum stress     (b) Displacement     (c) Rotation

Fig 17. Change in leg thickness

# 5. Optimization

Three optimization strategies were adopted for both parametric and numerical genetic algorithm optimizations in this section. The objective of the optimization strategy-1 is to minimize the mass while marginally reduce the maximum stress, displacement and rotation both at the jacket top and the mudline level (smaller than those of the original design). Since the ORIGH already adhered to the design codes' stipulations regarding ultimate and fatigue bearing capacity, the current optimization strategy is set aiming to reduce the mass of the jacket substructure while upholding the bearing capacity of the ORIGH. The second optimization strategy is a slight modification of the strategy-1, in which the constraints on the displacement and rotation at the mudline level are released. Instead, the rotation at the mudline level is expected to be smaller than 0.25° according to DNV code J101. Such simplification is assumed since the original design was conservative concerning the rotation at the mudline level whose value fall well below the requirement in the DNV code J101 for the ultimate and fatigue bearing capacity design. Comparing to the optimization strategy 1 and 2, the third strategy is the most aggressive one. The prescribed design constraints pertaining to rotational and displacement parameters at both the top of the jacket structure and the mudline level are delineated within the normative frameworks of ASCE-7-16 [31], Eurocode 7 [32], and DNV code J101. Within the ASCE-7-16, a displacement threshold of 0.2% relative to the structural height is stipulated for the uppermost portion of the jacket, resulting in a calculated value of 172 mm for the present configuration. Eurocode 7 specifies a maximal rotational displacement of 1/150 radians (equivalent to 0.38°) at the summit of the jacket for scenarios corresponding to the ultimate limit state. Conversely, the rotational limitation of 0.25° at the mudline level is prescribed by the DNV code J101. Notably, the precise vertical displacement limit at the mudline level is unspecified within the aforementioned design codes, thus no constraint is prescribed. Furthermore, it is incumbent that the stress levels across all structural constituents remain below the threshold of 355 MPa.

Two optimization methods are introduced and compared in this section, i.e., parametric optimization and numerical genetic algorithm optimization. The former was conducted based on the results of parametric studies, and further modified empirically according to the simulation results, while the numerical GA optimization method employed a population-based approach, iteratively evolving candidate solutions through selection, crossover, and mutation operations guided by a fitness function, mimicking natural selection to find approximate solutions to optimization problems.

## 5.1 Parametric optimization

As concluded from the parametric studies, the change in response per mass increment is shown in Table 5, and it will serve as the reference for the current parametric optimization. The change in the LT contributes most to the change in stress, the BW contribute most to the maximum displacement, the LD contribute most to the mudline displacement. The maximum rotation is influence most by the BW, whilst the mudline rotation is influenced mostly by the BD.

Table 5. Change in response per mass increment

| Parameter | Stress (MPa/ton) | Max Disp (mm/ton) | Mudline Disp (mm/ton) | Max rotation (deg/ton) | Mudline rotation (deg/ton) |
|---|---|---|---|---|---|
| BW | -0.246 | -0.403 | -0.010 | -3.62e-4 | +1.44e-4 |
| BD | -0.011 | -0.032 | -0.009 | -1.50e-4 | -2.34e-4 |
| BT | -0.193 | -0.070 | -0.003 | -1.88e-4 | -8.16e-5 |
| LD | -0.199 | -0.124 | -0.014 | -2.90e-4 | -2.83e-5 |
| LT | -0.262 | -0.082 | -0.00326 | -1.50e-4 | +3.02e-6 |

### 5.1.1 Parametric optimization: Strategy-1

To achieve the objectives of optimization strategy 1, a two-step optimization method was implemented. In the first step, the pile length was set to be 1400 mm, resulting in stress, displacement, and rotation reductions. The parameter combinations of BW, BD, BT, LD, and LT are shown in Table 6. For each combination chosen, the estimated contribution to the mass, maximum stress, maximum displacement and rotation, and mudline rotation were calculated based on factors in Table 5. For BW, since its increase exerts a substantial influence on the reduction of all the constraint indexes except the mudline rotation, the maximum value (40000 mm-Comb.6) was used. The enhancement of the mudline rotation due to the increase of BW was counteracted by rising BD as the increase in BD contributes significantly to reducing rotation, both at the top of the structure and at the mudline level. Therefore, the combination with maximum parametric values (Comb.6) was adopted. Increasing BT mainly contributes to the reduction of overall stress and rotation before up to Comb.3, after which the gradient of reduction became mild. Therefore, Comb.3 was used for BT to reduce the mass while maintain the structural response closely aligned with that of the ORIGH. The decision process of determining the parameter combination of LD was similar to that of BW, Comb.6 was applied. Increasing LT only has significant impact on the overall stress, Therefore, Comb.3 was used to slightly reduce the mass while keeping a structural response similar to that of the ORIGH.

Table 6. Selection of parameter combinations (Strategy 1)

|  | BW | BD | BT | LD | LT | Estimated Change | Simulated Change | Step-2 Simulation |
|---|---|---|---|---|---|---|---|---|
| Combination selected | Comb.6 | Comb.6 | Comb.3 | Comb. 6 | Comb.3 | - | - | - |
| Mass (tons) | +57 | +43 | -175 | +85 | -131 | -121 | -153 | 1671 |
| Max. stress (MPa) | -14.02 | -0.47 | 33.78 | -16.92 | +34.32 | +36.69 | +14.20 | 275.35 |
| Max displacement (mm) | -22.97 | -1.38 | 12.25 | -10.54 | +10.74 | -11.90 | -9.28 | 75.41 |
| Max rotation (°) | -2.06E-02 | -6.45E-03 | +3.29E-02 | -2.47E-02 | +1.97E-02 | +0.00082 | -0.0025 | 0.3522° |
| Mudline rotation (°) | 8.21E-03 | -1.01E-02 | +1.43E-02 | -2.41E-03 | -3.96E-04 | +0.00962 | -0.0086 | 0.1251° |

The simulation occupying the parameter combinations in Table 6 resulted in changes to the constraint indexes, as listed in the last column of Table 6. The constraint in the maximum stress was not fulfilled. Noting that the maximum stress occurred at the jacket top, so the second step optimization was conducted to empirically modify the parameters at level-3 and level-4 (as shown in Fig 2). Parameters in Comb.4 were used for BT and LT of members at level-3 and 4. All the constraints were satisfied with such modification, and the simulated results are presented in the last column of Table 6 and will be analysed in section 5.3.

### 5.1.2 Parametric optimization: Strategy-2

With respect to the optimization strategy-2, less attention was paid to the rotation at the mudline level, with a substitutional constraint being to ensure its value remained below 0.25°. In the context of the step-1 optimization, compared to the parameter selection in Table 6, Comb.4 was used for BD as the increase in mudline rotation caused by the variation of other parameters were allowed. The selected parameter combinations, the corresponding estimated changes as well as the simulated changes are outlined in Table 7. The constraints in the maximum stress and overall rotation were not satisfied. Similar to the step-2 optimization with respect to strategy-1, adjustments were made to BD and LT of members at level-3 and 4, that parameters in Comb.4 were employed. With such modifications the constraints were satisfied. Also, the simulation results of the optimized structure are presented in the last column of Table 7 and will be analysed in section 5.3.

Table 7. Selection of parameter combinations (Strategy 2)

|  | BW | BD | BT | LD | LT | Estimated Change | Simulated Change | Step-2 Simulation |
|---|---|---|---|---|---|---|---|---|
| Combination selected | Comb.6 | Comb.4 | Comb.3 | Comb. 6 | Comb.3 | - | - | - |
| Mass (tons) | +57 | -43 | -175 | +85 | -131 | -207 | -223 | 1603 |
| Max. stress (MPa) | -14.02 | +0.47 | +33.78 | -16.92 | +34.32 | +37.63 | +30.20 | 275.91 |
| Max displacement (mm) | -22.97 | +1.38 | 12.25 | -10.54 | +10.74 | -9.14 | -3.14 | 80.23 |
| Max rotation (°) | -2.06E-02 | +6.45E-03 | +3.29E-02 | -2.47E-02 | +1.97E-02 | +0.0137 | +0.0132 | 0.3637° |
| Mudline rotation (°) | 8.21E-03 | +1.01E-02 | +1.43E-02 | -2.41E-03 | -3.96E-04 | +0.0297 | +0.0301 | 0.1567° |

### 5.1.3 Parametric optimization: Strategy-3

In the context of optimization strategy-3, the parameters of optimization step-1 were selected with respect to the constraints in design codes. The parameter combinations selected were outlined in Table 8. Comparing to those listed in Table 7, Comb.4 was chosen for BW and Comb.5 was selected for LD allowing a significant increase in the maximum stress and displacement. The simulated results show that only the maximum rotation was above the code limit. Since the maximum rotation was at the top of the structure, the step-2 modification was conducted to change the diameters of members CS3 to VJ4 to Comb-6 and the BT at level-4 to Comb.6. The final simulated results are listed in the last column of Table 8.

Table 8. Selection of parameter combinations (Strategy 3)

|  | BW | BD | BT | LD | LT | Estimated Value | Simulated Value | Step-2 Simulation |
|---|---|---|---|---|---|---|---|---|
| Combination selected | Comb.4 | Comb.4 | Comb.3 | Comb. 5 | Comb.3 | - | - | - |
| Mass (tons) | -26 | -43 | -175 | 0 | -131 | 1406 | 1421 | 1517 |
| Max. stress (MPa) | +6.396 | +0.473 | +33.78 | 0 | +34.322 | 363.67 | 338.69 | 322.37 |
| Max displacement (mm) | +10.478 | +1.376 | +12.25 | 0 | +10.742 | 122 | 109.80 | 100.00 |
| Max rotation (°) | +9.41e-03 | +6.45E-03 | +3.29E-02 | 0 | +1.97E-02 | 0.4349 | 0.4152 | 0.3802° |
| Mudline rotation (°) | -3.74e-03 | +1.01E-02 | +1.43E-02 | 0 | -3.96E-04 | 0.1544 | 0.1357 | 0.1448° |

The results of the parametric optimization are summarized in Table 12.

### 5.2 Numerical Genetic Algorithm optimization

The parametric optimizations introduced in section 5.1 were conducted based on the simplification that parameters belonging to a single category underwent simultaneous variation during the optimization process. However, the impact arising from various permutations of parameter combinations within the same category on the resultant optimization outcomes was not taken into account. Consequently, the attainment of the most optimal parameter combination might not have been fully realized. Besides, the determination of the optimized design contingent on different design requirements relied on an empirical approach, potentially limiting its applicability within the realm of engineering design practices. As an alternative, a numerical GA optimization programme is proposed in this section with the FEM module to be called to solve the objective and constraint functions. The structure and the workflow of the numerical GA optimization programme are shown in Fig 18.

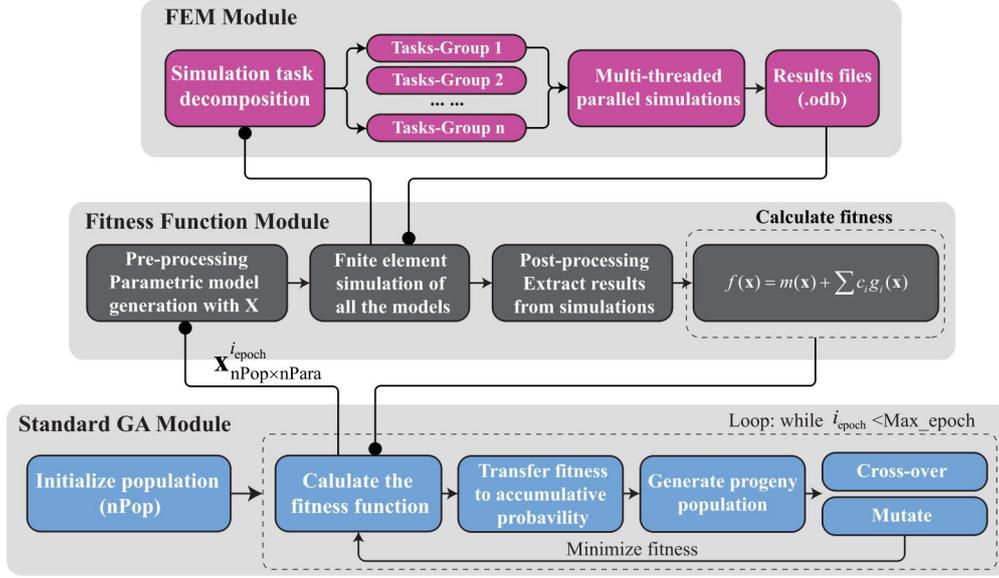

Fig 18. The structure and workflow of the numerical GA optimization programme

The entire programme consists three main modules, namely the standard GA module, the fitness function module and the finite element modelling module. The standard GA module was built following the traditional GA structure, with special attention being paid to the fitness calculation part leading to the fitness function module. Model parameters were passed into the fitness function module for parametric model generation first. All the generated modelled were passed into the FEM module for simulation, and the post-processing part was called to extract results and calculated the fitness values when all the result files (.odb) were ready. The models were imported to the FEM module in batch, and the batch size depends on the configuration of the computer being used. All the imported models were divided into several groups for multithreaded parallel simulations.

In numerical GA optimization, there is no need to make simplifications and assumptions on optimization parameters, so all the 48 parameters were involved. To ensure the optimized parameters are implementable in real engineering practices, all these parameters were discretely valued within the given ranges, and the limits and interval of all the parameters are listed in the Appendix Table-A 1.

To conduct optimization, the population size ( nPop ) was set to be 300, the crossover probability ( $P_c$ ) was 0.4. A rank based adaptive mutation probability algorithm was adopted to avoid premature convergence [33], in which the mutation probabilities ( $P_m$ ) were determined based on the chromosome rank ( $\Re$ ), and $\Re$ was decided based on the relative fitness in population of a specific chromosome. The expression for $P_m$ writes:

$$P_m = P_m^{\max}\left(1 - \frac{\Re - 1}{\text{nPop} - 1}\right) \tag{14}$$

where $P_m^{\max} = 0.8$ is the maximum mutation probability. The optimization iteration terminates when the minimum fitness value remains unchanged over 20 generations.

To reduce the mass of the jacket substructure, three optimization strategies subjected to three groups of constraints (as will be outlined in Eqs., and ) will be performed in this section. The optimizations including all the subsequential simulations were performed on a personal computer with Intel(R) Core(TM) i9-13900k CPU @ 5.8 GHz, 64 GB RAM, and Ubuntu 23.04 OS.

### 5.2.1 Numerical GA optimization: Strategy-1

With respect to the optimization strategy-1, the optimization problem can be expressed as:

**Minimize:** $m(\mathbf{X})$

**Subjected to:**

$$\begin{aligned}
&|\sigma(\mathbf{X})|_{max} < |\sigma(\mathbf{X}_0)|_{max} \rightarrow g_1(\mathbf{X}) = \max(0, |\sigma(\mathbf{X})|_{max} - |\sigma(\mathbf{X}_0)|_{max}) \\
&|\mathbf{u}_t(\mathbf{X})|_{max} < |\mathbf{u}_t(\mathbf{X}_0)|_{max} \rightarrow g_2(\mathbf{X}) = \max(0, |\mathbf{u}_t(\mathbf{X})|_{max} - |\mathbf{u}_t(\mathbf{X}_0)|_{max}) \\
&|\phi_t(\mathbf{X})|_{max} < |\phi_t(\mathbf{X}_0)|_{max} \rightarrow g_3(\mathbf{X}) = \max(0, |\phi_t(\mathbf{X})|_{max} - |\phi_t(\mathbf{X}_0)|_{max}) \\
&|\mathbf{u}_m(\mathbf{X})|_{max} < |\mathbf{u}_m(\mathbf{X}_0)|_{max} \rightarrow g_4(\mathbf{X}) = \max(0, |\mathbf{u}_m(\mathbf{X})|_{max} - |\mathbf{u}_m(\mathbf{X}_0)|_{max}) \\
&|\phi_m(\mathbf{X})|_{max} < |\phi_m(\mathbf{X}_0)|_{max} \rightarrow g_5(\mathbf{X}) = \max(0, |\phi_m(\mathbf{X})|_{max} - |\phi_m(\mathbf{X}_0)|_{max})
\end{aligned} \quad (15)$$

where $\mathbf{X}$ is the geometric and sectional parameters, and $\mathbf{X}_0$ refers to the parameters of the original design. $m(\cdot)$ is the objective function referring to the overall mass of the structure, $\sigma_{max}$ is the overall maximum stress, $|\mathbf{u}_t(\cdot)|_{max}$ is the maximum displacement at the jacket top, $|\phi_t(\cdot)|_{max}$ is the maximum rotation at the jacket top about the z axis, $|\phi_m(\cdot)|_{max}$ is the maximum rotation at the mudline level about the z axis, and the vertical displacement. $g_i(\cdot)$ are the constraint functions.

The static penalty function is adopted to build the fitness function with constraints,

$$f(\mathbf{X}) = m(\mathbf{X}) + \sum c_i g_i(\mathbf{X}) \quad (16)$$

where, $c_i$ are the penalty constants and were all set to be $10^5$ in current optimizations.

Three optimization attempts were conducted to ensure the results obtained was the optimal. Fig 19 shows the fitness landscape for optimization strategy 1, and Table 9 depicts the comparative statistical data of the three attempts. It can be found in both Fig 19 and Table 9 that the best optimization was achieved by attempt-1, as evidenced by the fitness converging to $f(\mathbf{x}) = 1482$, which leads to a mass reduction of 16.8% (299 ton). For all the three attempts, the simulated maximum stress, displacement and rotation were all reduced marginally. The best fitness values exhibit clear downward trends before ~60 generations, and tended to decline gently until converge. The results of optimization attempt-1 will be used in this section for further analysis. The optimized parameters of attempt-1 (named GA-St1-Apt1) are specifically given in the Appendix Table-A 3.

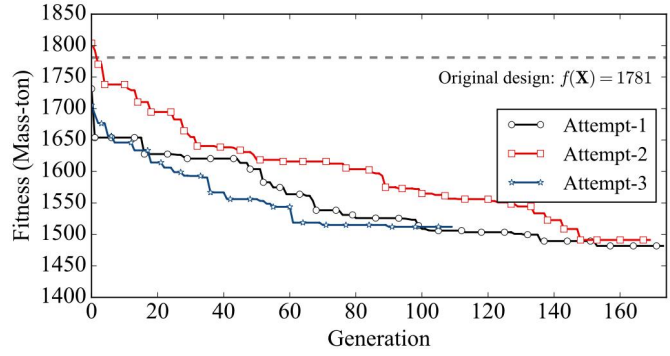

Fig 19. Fitness landscape strategy-1

Table 9. Comparative data of the three attempts of optimization strategy-1

| Optimization strategy-1 | Attempt-1 | Attempt-2 | Attempt-3 | ORIGH (Limits) |
|---|---|---|---|---|
| Optimum achieved (ton) | **1482(best)** | 1491 | 1512 | 1781 |
| No. of generations | 174 | 170 | 110 | - |
| Max. stress (MPa) | 267.4 | 262.0 | 266.7 | 288.7 |
| Max. top displacement (mm) | 86.71 | 84.43 | 85.14 | 87.12 |
| Max. top rotation (degree) | 0.292 | 0.292 | 0.290 | 0.309 |
| Max. mudline displacement (mm) | 17.43 | 17.33 | 17.46 | 17.56 |
| Max. mudline rotation (degree) | 0.116 | 0.120 | 0.118 | 0.120 |

Fig 20 illustrates the parameter fluctuations during the optimization process in attempt-1, with each parameter's variation was quantified as $\sigma_i = (X_i - X_{i,0})/X_{i,0}$. There are three sensitive segments (including 11 parameters) over the 48 genetic parameters including the base width, diameter and thickness of members SB1, SB1D, BJ2, BJ3 (diameter only), BJ4, and UPC3 (diameter only). This trend suggests a heightened sensitivity of these member dimensions to the optimization objective, while subjecting to the constraints outlined by Eq.. In addition, it is found that reducing the thickness of members BC2, UPC, CS1, CS2, SB4, UPC3, and BC4, along with decreasing the diameter of members CS3 and BJ4, contribute significantly to achieve the optimization objective ($\sigma_i \leq -0.3$).

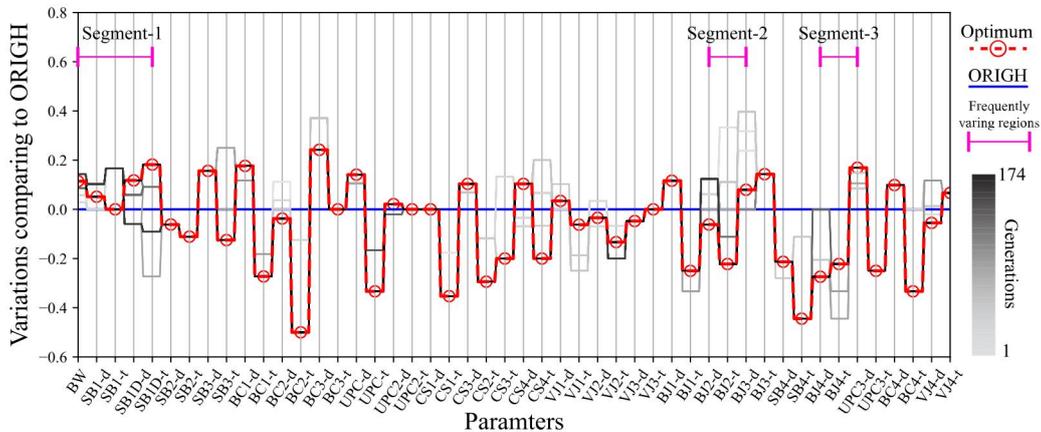

Fig 20. Parameter sensitivity analysis (Strategy-1, attempt-1)

### 5.2.2 Numerical GA optimization: Strategy-2

With respect to the optimization strategy-2, the optimization problem can be written as:

**Minimize:** $\quad m(\mathbf{X})$

**Subjected to:**
$$|\sigma(\mathbf{X})|_{\max} < |\sigma(\mathbf{X}_0)|_{\max} \rightarrow g_1(\mathbf{X}) = \max(0, |\sigma(\mathbf{X})|_{\max} - |\sigma(\mathbf{X}_0)|_{\max})$$
$$|\mathbf{u}_t(\mathbf{X})|_{\max} < |\mathbf{u}_t(\mathbf{X}_0)|_{\max} \rightarrow g_2(\mathbf{X}) = \max(0, |\mathbf{u}_t(\mathbf{X})|_{\max} - |\mathbf{u}_t(\mathbf{X}_0)|_{\max}) \quad (17)$$
$$|\phi_t(\mathbf{X})|_{\max} < |\phi_t(\mathbf{X}_0)|_{\max} \rightarrow g_3(\mathbf{X}) = \max(0, |\phi_t(\mathbf{X})|_{\max} - |\phi_t(\mathbf{X}_0)|_{\max})$$
$$|\phi_m(\mathbf{X})|_{\max} < 0.25° \rightarrow g_4(\mathbf{X}) = \max(0, |\phi_m(\mathbf{X})|_{\max} - 0.25°)$$

In Eq., the displacement at the mudline level was not constraint since it is not specified in the code. With Eq., the fitness function was constructed according to Eq.. Three optimization attempts were conducted to avoid a premature result. Fig 21 shows the fitness landscape for optimization strategy 2, and

Table 10 depicts the comparative data of the three attempts. It is found that the optimums achieved by the three attempts were different but all close to 1400 ton, and all the constraints outlined in Eq. were satisfied. The variation in optimum values of the three attempts can be attributed to the stochastic nature of GA. The best result was obtained by attempt-2 (GA-St2-Apt2) with $f(\mathbf{X}) = 1384$, leading to a mass reduction of 22.2% (397 ton). The optimization process of attempt-2 will be analysed in this section, and the optimized parameters are listed in the Appendix Table-A 3.

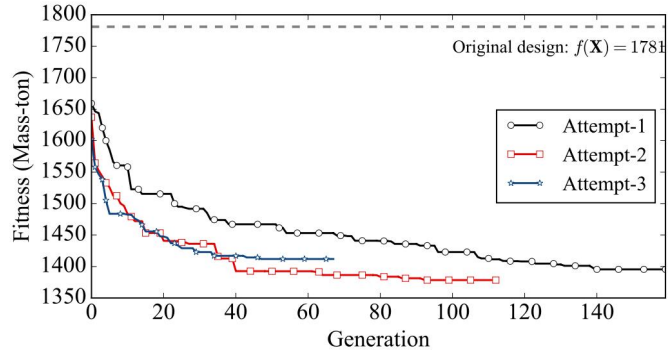

Fig 21. Fitness landscape for strategy-2

Table 10. Comparative data of the three attempts of optimization strategy-2

| Optimization strategy-2 | Attempt-1 | Attempt-2 | Attempt-3 | ORIGH | Limits |
|---|---|---|---|---|---|
| Optimum achieved (ton) | 1396 | **1384** (best) | 1412 | 1781 | <1781 |
| No. of generations | 161 | 113 | 78 | - | - |
| Max. stress (MPa) | 273.5 | 281.7 | 270.5 | 288.7 | <288.7 |
| Max. top displacement (mm) | 86.92 | 87.01 | 87.05 | 87.12 | <87.12 |
| Max. top rotation (degree) | 0.302 | 0.301 | 0.296 | 0.309 | <0.309 |
| Max. mudline rotation (degree) | 0.155 | 0.158 | 0.156 | 0.120 | <0.250 |

Fig 22 illustrates the parameter fluctuations during the optimization process in attempt-2. The analysis reveals two distinct segments wherein genetic parameters exhibited notable variability throughout the optimization process. Notably, parameters encompassing the diameters and thicknesses of members SB1D, SB2, SB3, CS2, CS3, CS4, VJ1, and VJ2 (including 13 parameters in total) exhibit particularly pronounced variations. This pattern indicates a strong sensitivity of these member dimensions to the optimization objective, while concurrently subjecting to the constraints specified by Eq.. In addition, it is found that decrease in the thickness of BC1, BC2, BC4, CS1, and CS2 contributed significantly to the reduction of the jacket's mass ($\sigma_i \leq -0.3$). The thickness of BJ3 increases by 42%, which is opposite to the optimization objective but may be beneficial to satisfying specific constraints.

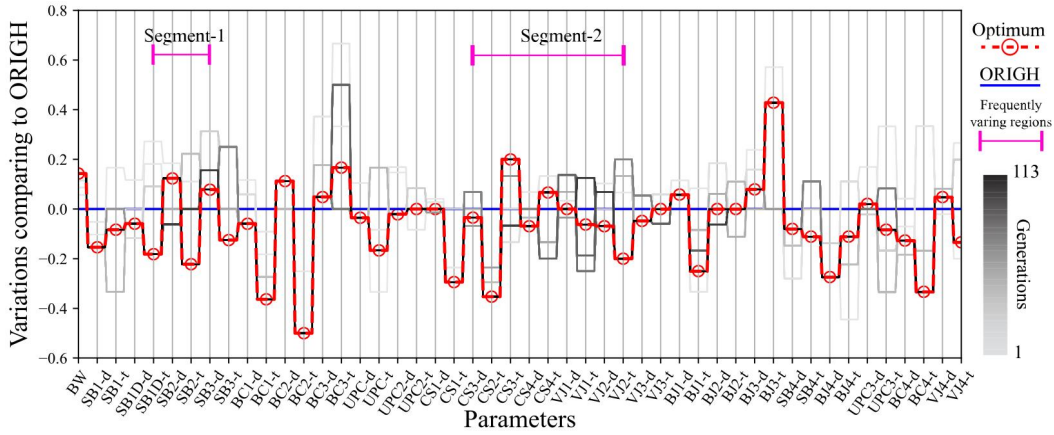

Fig 22. Parameter sensitivity analysis (Strategy-2, attempt-2)

### 5.2.3 Numerical GA optimization: Strategy-3

The collective assertions elucidated above furnish the foundational premises underpinning the subsequent formulation of the optimization problem, as follows:

**Minimize:**   $m(\mathbf{X})$

**Subjected to:**  $\sigma_{\max}(\mathbf{X}) < \sigma_y = 355$ MPa $\rightarrow g_1(\mathbf{X}) = \max(0, \sigma_{\max}(\mathbf{X}) - 355)$

$|\mathbf{u}_t(\mathbf{x})|_{\max} < 172$ mm $\rightarrow g_2(\mathbf{x}) = \max(0, |\mathbf{u}_t(\mathbf{x})|_{\max} - 172)$   (18)

$|\phi_t(\mathbf{X})|_{\max} < 0.3819° \rightarrow g_3(\mathbf{X}) = \max(0, |\phi_t(\mathbf{X})|_{\max} - 0.3819°)$

$|\phi_m(\mathbf{X})|_{\max} < 0.25° \rightarrow g_5(\mathbf{X}) = \max(0, |\phi_m(\mathbf{X})|_{\max} - 0.25°)$

Three optimization attempts were conducted to ensure the results obtained is the optimal. Fig 23 shows the fitness simulation graph for optimization strategy 3, and Table 11 depicts the comparative statistical data of the three attempts. It can be seen that the optimums achieved by the three attempts are close, but the attempt-2 underwent the largest generations and achieved the smallest fitness value as $f(\mathbf{x}) = 1171$ leading to a mass reduction of 34.3% (610 ton). The optimized parameters of attempt-2 (named GA-St3-Apt2) are listed in the Appendix Table-A 3, and the optimization of attempt-2 will be analysed in this section.

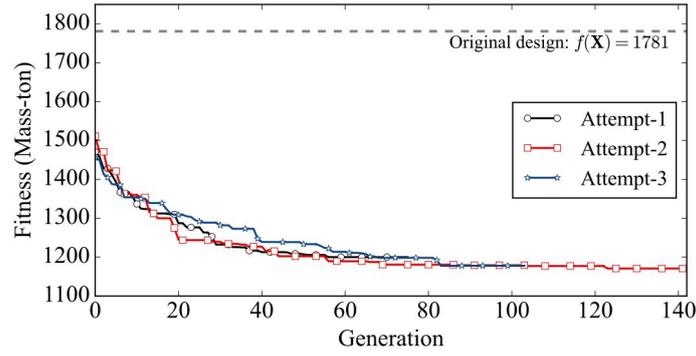

Fig 23. Fitness landscape strategy-3

Table 11. Comparative data of the three attempts of optimization strategy-3

| Optimization strategy-3 | Attempt-1 | Attempt-2 | Attempt-3 | ORIGH | Limits |
|---|---|---|---|---|---|
| Optimum achieved | 1200 | **1171 (best)** | 1178 | 1781 | - |
| No. of generations | 76 | 143 | 104 | - | - |
| Max. stress (MPa) | 321.2 | 332.3 | 327.0 | 288.7 | <355.0 |
| Max. top displacement (mm) | 156.53 | 156.21 | 168.59 | 87.12 | <172.00 |
| Max. top rotation (degree) | 0.331 | 0.324 | 0.329 | 0.309 | <0.380 |
| Max. mudline rotation (degree) | 0.118 | 0.136 | 0.121 | 0.120 | <0.250 |
| Max. mudline displacement (mm) | 19.96 | 20.27 | 20.19 | 17.56 | - |

Fig 24 shows the variations of the optimization parameters in attempt-2, and three distinct segments including 25 frequently varying parameters are found. The 25 parameters contain the based width, diameter and thickness of SB1, SB1D, SB2, SB3 (diameter only), CS3 (thickness only), CS4, VJ1, VJ2, VJ3 (diameter only), BJ1 (thickness only), BJ2, BJ3, BJ4 and SB4, which indicates the structure is sensitive to these parameters while achieving the optimization goal. Most of the parameters were reduced after the optimal solution was achieved, especially the thickness of members SB1, BC1, BC2, UPC, CS1, CS2, SB4, and BJ4 ($\sigma_i \leq -0.3$).

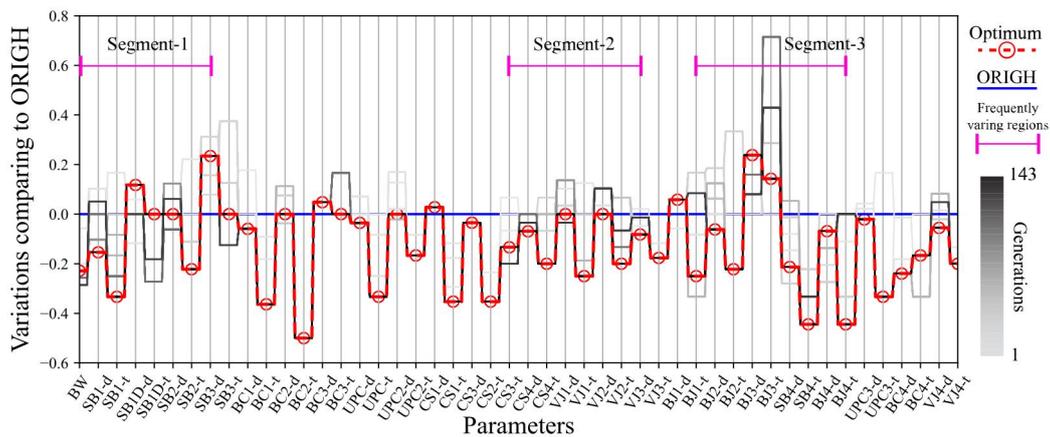

Fig 24. Parameter sensitivity analysis (Strategy-3, attempt-2)

## 5.3 Comparison of the optimization results

The optimized jacket structures obtained through parametric optimizations (PARM-Opt1, PARM-Opt2 and PARM-Opt3) and numerical GA optimizations (GA-St1-Apt1, GA-St2-Apt2 and GA-St3-Apt2) are comprehensively compared in this section to analyse the difference of the two optimizations methods and the three optimization strategies. The simulated results are specifically outlined in Table 12, and the stress, displacement and rotation contours of the optimal configurations of the 6 optimizations and the ORIGH are shown in Fig 25.

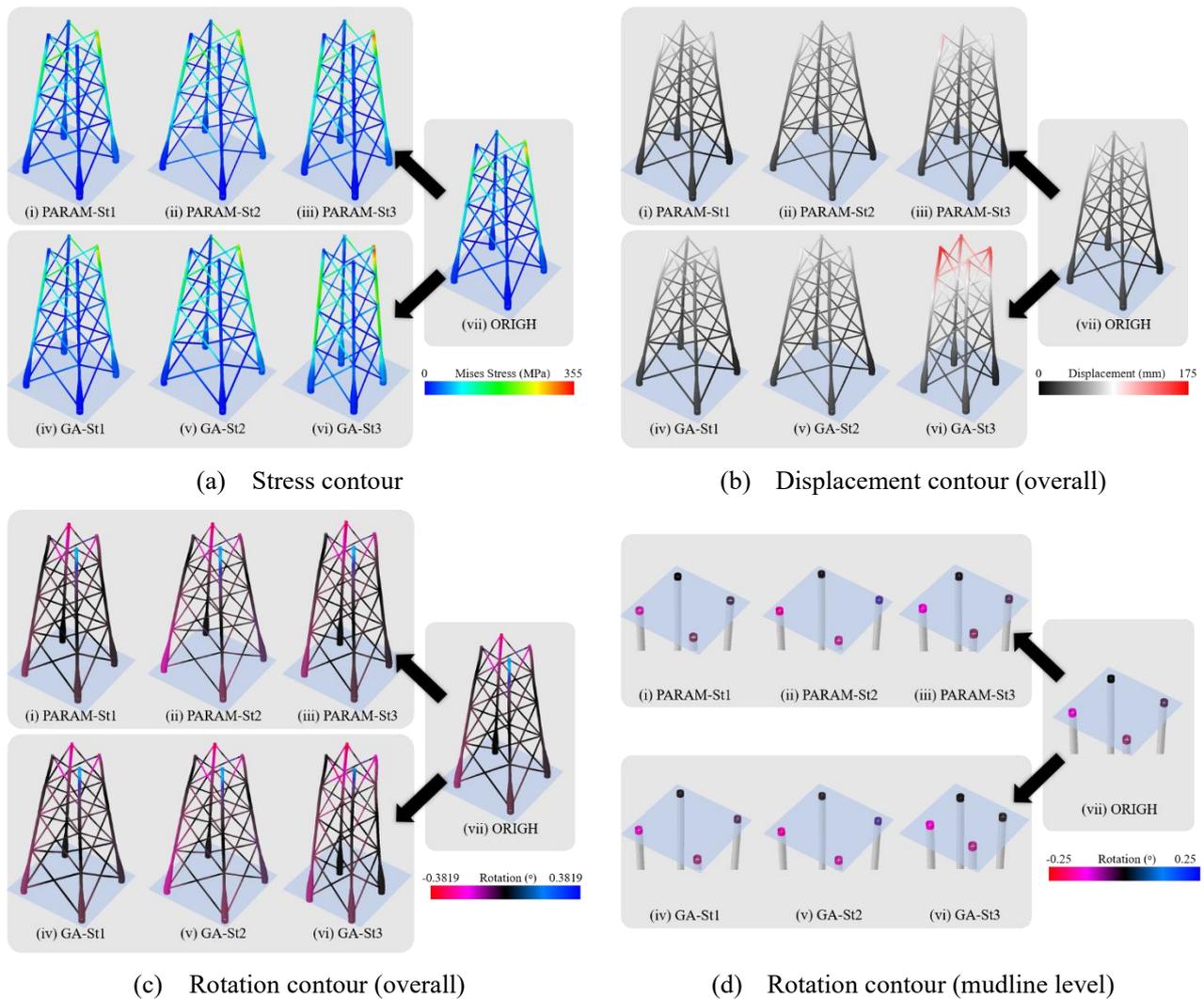

(a) Stress contour  (b) Displacement contour (overall)

(c) Rotation contour (overall)  (d) Rotation contour (mudline level)

Fig 25. Comparison of stress, displacement and rotation contours simulated using different optimization methods and strategies

Table 12. Comparison of optimization results

| Model | Mass (ton) | Max. Stress (MPa) | Max. Displacement (mm) | Max Rotation (°) | Max. Rotation at Mudline (°) |
|---|---|---|---|---|---|
| PARAM-St1 | 1671 | 275.4 | 75.41 | 0.3522 | 0.1251 |
| PARAM-St2 | 1603 | 275.9 | 80.23 | 0.3637 | 0.1567 |
| PARAM-St3 | 1517 | 322.4 | 100.01 | 0.3802 | 0.1448 |
| GA-St1 | 1482 | 267.4 | 86.72 | 0.2921 | 0.1163 |
| GA-St2 | 1384 | 281.7 | 87.01 | 0.3013 | 0.1582 |
| GA-St3 | 1171 | 332.3 | 156.21 | 0.3242 | 0.1360 |
| ORIGH | 1781 | 288.7 | 87.15 | 0.3665 | 0.1342 |
| Code limits | - | 355.0 | 175.00 | 0.3819 | 0.2500 |

For the optimization problem in this paper, the goal was to find or approach the global optimal parameter combination in a complex, and noisy search space. Owning to the inherent flexibility of GA optimization, each parameter can be optimized individually without relying on gradient (trend) analysis which was crucial for parametric optimization. Also, the simulated stress, displacement and rotation of GA optimized models were more approaching to the limits of the three optimization strategies, as shown in Fig 25 and Table 12. As the result, GA optimizations were more aggressive, and the comparative results of the two optimization methods show that the optimal jacket masses achieved by numerical GA optimizations were all smaller than those obtained through parametric optimizations. However, though an efficient jacket FEM model was developed in this work, GA optimization was still more computational expensive than parametric optimization. The selection between the two optimization methods in practical engineering design necessitates finding a balance between computational efficiency and optimization outcomes.

The simulated results in Fig 25 and Table 12 also reflect the difference of the three optimization strategies. The results of strategy-1 both from parametric and GA optimizations were closely aligned to those of ORIGH, which indicates that the optimization processes successfully adhered to the specified problem constraints. Strategy-1 is the most conservative one comparing to the rest two and is the mostly preferred in real engineering design practices because all the design indexes fall below those of ORIGH. For strategy-2, the common change of the two optimization methods was to increase the base width, which allows a larger rotation at the lower part of the jacket structure. This optimization strategy may be favoured when the fatigue capacity of the grout connection is deemed adequate. Strategy-3 is the most aggressive of the three. Given that the maximum displacement defined in the design code was considerably larger than that of ORIGH, the optimizations in this strategy were geared towards a significant increase in displacement at the jacket tops, as illustrated in Fig 25 (b). Consequently, this led to a substantial increase in stress at the top, as shown in Fig 25(a). Adopting strategy-3 in practical engineering design requires a further analysis of the fatigue capacity of the jacket structure, and modifications based on the analysis results are compulsory as well.

## 6. Conclusions

An efficient finite element model for simulating the behaviour of a jacket substructure under the combined self-weight, wind, and wave loads was developed in this paper. This model also takes into account the realistic interactions between the structure and the surrounding soil. The FEM model was further used as the analysis tool for parametric optimization and solver for GA optimization. The results show that both the two optimization methods can efficiently reduce the mass of the jacket structure under three optimization strategies while fulfilling all the design requirements. The application scenarios and suggestions of two optimization methods and three optimization strategies were also given at the end of the paper after comprehensive comparisons. The main conclusions are:

1) The developed FEM jacket model could be built with parametric description so that models with different geometries and sections can be constructed in batches. The FEM model integrated a DLOAD subroutine for modelling the water depth, location and sectional dimension related wave load and a UEL subroutine for modelling the nonlinear soil-pile interactions. The proposed model showed its efficiency in model generation and structural simulation, so that it can be adopted in optimizations of jacket substructures.

2) Parametric analysis and optimization were conducted with the aid of the proposed FEM. The trends of the mass and response of the jacket substructure with respect to the variation of geometric and sectional parameters were studied in depth, which guided the selection of parameter combinations in parametric optimizations. The mass reductions achieved with parametric optimization were 6.2%, 10% and 14.8% for the three strategies respectively.

3) A numerical GA optimization was carried out with the proposed FEM embedded in a rank-based GA framework acting as the solver to optimization objective and constraint functions. The mass of the jacket substructure reduced by 16.8%, 22.3% and 34.3% for the three strategies respectively, indicating that the numerical GA optimization was more aggressive in terms of the mass reduction comparing to the parametric optimization.

## Acknowledgements

The authors gratefully acknowledge financial support from China Postdoctoral Science Foundation (No. 2022M722792), Zhejiang Provincial Natural Science Foundation of China (HZ19E090002) and National Natural Science Foundation of China (Nos. 51779224 and 51579221)

## Appendices

Table-A 1. Limits and intervals of parameters used in GA optimizations (unit: mm)

| Label | Lower limit | Upper limit | Interval | Label | Lower limit | Upper limit | Interval |
|---|---|---|---|---|---|---|---|
| SB1-d | 825 | 1075 | 50 | SB1-t | 40 | 75 | 5 |
| SB1D-d | 750 | 950 | 50 | SB1D-t | 40 | 75 | 5 |
| SB2-d | 760 | 960 | 50 | SB2-t | 36 | 60 | 5 |
| SB3-d | 690 | 940 | 50 | SB3-t | 35 | 60 | 5 |
| BC1-d | 800 | 1000 | 50 | BC1-t | 35 | 70 | 5 |
| BC2-d | 770 | 890 | 30 | BC2-t | 20 | 50 | 5 |
| BC3-d | 650 | 850 | 40 | BC3-t | 30 | 50 | 5 |
| UPC-d | 1350 | 1650 | 50 | UPC-t | 40 | 70 | 5 |
| UPC2-d | 1350 | 1650 | 30 | UPC2-t | 40 | 70 | 5 |
| CS1-d | 3550 | 3800 | 50 | CS1-t | 55 | 85 | 5 |
| CS3-d | 1350 | 1600 | 50 | CS2-t | 55 | 90 | 5 |
| CS3-t | 60 | 90 | 5 | CS4-d | 1350 | 1600 | 50 |
| CS4-t | 60 | 90 | 5 | VJ1-d | 1410 | 1660 | 50 |
| VJ1-t | 60 | 90 | 5 | VJ2-d | 1350 | 1600 | 50 |
| VJ2-t | 60 | 90 | 5 | VJ3-d | 1350 | 1600 | 50 |
| VJ3-t | 60 | 90 | 5 | BJ1-d | 810 | 960 | 50 |
| BJ1-t | 40 | 75 | 5 | BJ2-d | 760 | 960 | 50 |
| BJ2-t | 35 | 60 | 5 | BJ3-d | 630 | 880 | 50 |
| BJ3-t | 35 | 60 | 5 | SB4-d | 540 | 790 | 50 |
| SB4-t | 25 | 50 | 5 | BJ4-d | 530 | 780 | 50 |
| BJ4-t | 25 | 50 | 5 | UPC3-d | 1390 | 1650 | 30 |
| UPC3-t | 40 | 80 | 5 | BC4-d | 540 | 820 | 40 |
| BC4-t | 20 | 50 | 5 | VJ4-d | 1370 | 1620 | 50 |
| VJ4-t | 60 | 95 | 5 | BW | 25000 | 40000 | 1000 |

Table-A 2. Optimized parameters (Parametric optimization), unit: mm

| | PARAM-st1 | | PARAM-st2 | | PARAM-st3 | |
|---|---|---|---|---|---|---|
| Section Label | Diameter | Thickness | Diameter | Thickness | Diameter | Thickness |
| V1 | 4200 | 65 | 4200 | 65 | 4200 | 65 |
| V2 | 3600 | 85 | 3600 | 85 | 3600 | 85 |
| CS1 | 3750 | 75 | 3750 | 75 | 3600 | 75 |
| CS2 | 3750-1600 | 75 | 3750-1600 | 75 | 3600-1600 | 75 |
| CS3 | 1600 | 65 | 1600 | 65 | 1600 | 65 |
| CS4 | 1600 | 65 | 1600 | 65 | 1600 | 65 |
| VJ1 | 1610 | 70 | 1610 | 70 | 1610 | 70 |
| VJ2 | 1600 | 65 | 1600 | 65 | 1600 | 65 |
| VJ3 | 1600 | 65 | 1600 | 65 | 1600 | 65 |
| VJ4 | 1620 | 75 | 1620 | 75 | 1620 | 75 |

| Section Label | | | | | |
|---|---|---|---|---|---|
| UPC | 1570 | 50 | 1570 | 50 | 1570 | 50 |
| UPC2 | 1570 | 50 | 1570 | 50 | 1570 | 50 |
| UPC3 | 1570 | 50 | 1570 | 55 | 1570 | 50 |
| SB1 | 1015 | 50 | 935 | 50 | 935 | 50 |
| SB1D | 890 | 45 | 810 | 45 | 810 | 45 |
| SB2 | 850 | 35 | 770 | 35 | 770 | 35 |
| SB3 | 780 | 40 | 700 | 40 | 700 | 35 |
| SB4 | 680 | 35 | 640 | 35 | 600 | 45 |
| BC1 | 890 | 45 | 810 | 45 | 810 | 45 |
| BC2 | 840 | 30 | 760 | 30 | 760 | 30 |
| BC3 | 750 | 25 | 670 | 25 | 670 | 20 |
| BC4 | 660 | 25 | 620 | 25 | 580 | 35 |
| BJ1 | 900 | 50 | 820 | 50 | 820 | 50 |
| BJ2 | 850 | 35 | 770 | 35 | 770 | 35 |
| BJ3 | 770 | 40 | 690 | 40 | 690 | 35 |
| BJ4 | 670 | 30 | 630 | 30 | 590 | 40 |
| Base width | 35000 | | 40000 | | 32500 | |

Table-A 3. Optimized parameters (Numerical GA optimization), unit: mm

| | GA-st1 | | GA-st2 | | GA-st3 | |
|---|---|---|---|---|---|---|
| Section Label | Diameter | Thickness | Diameter | Thickness | Diameter | Thickness |
| V1 | 4200 | 65 | 4200 | 65 | 4200 | 65 |
| V2 | 3600 | 85 | 3600 | 85 | 3600 | 85 |
| CS1 | 3600 | 55 | 3650 | 55 | 3700 | 55 |
| CS2 | 3600-1600 | 60 | 3650-1550 | 55 | 3700-1400 | 55 |
| CS3 | 1600 | 60 | 1550 | 65 | 1400 | 65 |
| CS4 | 1600 | 60 | 1450 | 60 | 1350 | 60 |
| VJ1 | 1510 | 75 | 1660 | 65 | 1460 | 60 |
| VJ2 | 1400 | 65 | 1400 | 80 | 1450 | 60 |
| VJ3 | 1370 | 80 | 1620 | 60 | 1370 | 60 |
| VJ4 | 1400 | 85 | 1400 | 80 | 1350 | 70 |
| UPC | 1620 | 40 | 1620 | 60 | 1370 | 40 |
| UPC2 | 1660 | 45 | 1510 | 45 | 1390 | 40 |
| UPC3 | 1450 | 60 | 1420 | 60 | 1420 | 50 |
| SB1 | 1025 | 60 | 875 | 40 | 825 | 40 |
| SB1D | 950 | 65 | 800 | 40 | 950 | 55 |
| SB2 | 760 | 40 | 760 | 35 | 810 | 35 |
| SB3 | 590 | 25 | 690 | 40 | 590 | 25 |
| SB4 | 740 | 35 | 690 | 45 | 790 | 30 |
| BC1 | 1000 | 40 | 800 | 35 | 800 | 35 |
| BC2 | 770 | 20 | 800 | 20 | 800 | 20 |
| BC3 | 780 | 20 | 620 | 20 | 540 | 25 |
| BC4 | 770 | 30 | 730 | 40 | 650 | 30 |
| BJ1 | 960 | 45 | 810 | 60 | 910 | 45 |
| BJ2 | 760 | 35 | 960 | 40 | 760 | 35 |
| BJ3 | 530 | 35 | 580 | 40 | 680 | 25 |
| BJ4 | 680 | 40 | 630 | 60 | 780 | 40 |
| Base width | 39000 | | 38000 | | 27000 | |